\documentclass[a4paper,11pt]{article}
\usepackage{jheppub} 
\usepackage{bbm,bm,graphicx,mathtools,color,slashed}
\usepackage{amssymb,amsmath}
\usepackage{hyperref}
\usepackage[T1]{fontenc}

\hypersetup{
setpagesize=false,
 bookmarksnumbered=true,%
 bookmarksopen=true,%
 colorlinks=true
}

\usepackage[all]{xy}

\usepackage[framemethod=tikz]{mdframed}
\usepackage{tikz}

\usepackage[normalem]{ulem}  

\newcommand{\tr}{\mathop{\mathrm{tr}}}
\newcommand{\Tr}{\mathop{\mathrm{Tr}}}
\newcommand{\im}{\mathop{\mathrm{Im}}}

\newcommand{\diff}{\mathrm{d}} 
\newcommand{\rmi}{\mathrm{i}} 
\newcommand{\rme}{\mathrm{e}} 

\newcommand{\para}{\parallel} 
\newcommand{\etas}{\eta_s}

\newcommand{\hzero}{\hat{0}}
\newcommand{\hone}{\hat{1}}
\newcommand{\htwo}{\hat{2}}
\newcommand{\hthree}{\hat{3}}

\newcommand{\ha}{\hat{a}}
\newcommand{\hb}{\hat{b}}
\newcommand{\hc}{\hat{c}}
\newcommand{\hd}{\hat{d}}

\newcommand{\hy}{\hat{y}}

\newcommand{\ds}{\displaystyle}

\newcommand{\hH}{\hat{H}}

\newcommand{\tilT}{\widetilde{T}}

\newcommand{\hT}{\hat{T}}

\newcommand{\ptc}[1]{{\bar{#1}}}
\newcommand{\hSigma}{\hat{\Sigma}}

\newcommand{\hTheta}{\hat{\Theta}}

\newcommand{\hA}{\hat{A}}
\newcommand{\hB}{\hat{B}}

\newcommand{\hrhoeq}{\hat{\rho}_{\mathrm{eq}}}

\newcommand{\hOcal}{\hat{\mathcal{O}}}

\newcommand\Scal{\mathcal{S}}

\newcommand\Gcal{\mathcal{G}}
\newcommand\Lcal{\mathcal{L}}

\newcommand\Ocal{\mathcal{O}}

\newcommand\Rcal{\mathcal{R}}

\newcommand{\average}[1]{\langle#1\rangle}
\newcommand{\baverage}[1]{\big\langle#1\big\rangle}

\newcommand{\la}[1]{\overleftarrow{#1}}
\newcommand{\ra}[1]{\overrightarrow{#1}}

\newcommand\diag{\operatorname{diag}}

\newcommand\SU{{\rm SU}}

\newcommand{\bx}{\bm{x}}

\newcommand{\bk}{\bm{k}}
\newcommand{\bv}{\bm{v}}
\newcommand{\bnab}{\bm{\nabla}}
\newcommand{\bzero}{\bm{0}}

\newcommand{\tilep}{\widetilde{\epsilon}}

\newcommand{\tilG}{\widetilde{G}}
\newcommand{\tilpi}{\widetilde{\pi}}
\newcommand{\tilsigma}{\widetilde{\sigma}}
\newcommand{\tilSigma}{\widetilde{\Sigma}}
\newcommand{\tilTheta}{\widetilde{\Theta}}

\newcommand{\vep}{\varepsilon}

\newcommand{\U}{\text{U}}

\newcommand{\with}{\quad\mathrm{with}\quad}
\newcommand{\eqq}{~\Leftrightarrow~}
\newcommand{\qcd}{\mathrm{QCD}}

\newcommand{\eq}{\mathrm{eq}}
\newcommand{\ext}{\mathrm{ext}}
\newcommand{\ret}{\mathrm{R}}

\newcommand{\overc}[1]{{\mathring{#1}}}
\newcommand{\comega}{\overc{\omega}}
\newcommand{\cD}{\overc{D}}

\title{
Relativistic spin hydrodynamics with torsion \\
and linear response theory for spin relaxation
}

\author[a,b]{Masaru Hongo,}
\author[c,d]{Xu-Guang Huang,}
\author[e]{Matthias Kaminski,}
\author[a]{Mikhail Stephanov,}
\author[a]{Ho-Ung Yee}

\affiliation[a]{Department of Physics, University of Illinois, Chicago, IL 60607, USA}
\affiliation[b]{RIKEN iTHEMS, RIKEN, Wako 351-0198, Japan}
\affiliation[c]{Physics Department and Center for Field Theory and Particle Physics, Fudan University, Shanghai 200433, China}
\affiliation[d]{Key Laboratory of Nuclear Physics and Ion-beam Application (MOE), Fudan University, Shanghai 200433, China}

\affiliation[e]{Department of Physics and Astronomy, University of Alabama, Tuscaloosa, AL 35487, USA}

\emailAdd{hongo@uic.edu, huangxuguang@fudan.edu.cn, mski@ua.edu, misha@uic.edu, hyee@uic.edu
}

\abstract{
Using the second law of local thermodynamics and the first-order Palatini formalism, we formulate relativistic spin hydrodynamics for quantum field theories with Dirac fermions, such as QED and QCD, in a torsionful curved background.
We work in a regime where spin density, which is assumed to relax much slower than other non-hydrodynamic modes, is treated as an independent degree of freedom in an extended hydrodynamic description.
Spin hydrodynamics in our approach contains only three non-hydrodynamic modes corresponding to a spin vector, whose relaxation time is controlled by a new transport coefficient: the rotational viscosity. 
We study linear response theory and observe an interesting mode mixing phenomenon between the transverse shear and the spin density modes. 
We propose several field-theoretical ways to compute the spin relaxation time and the rotational viscosity, via the Green-Kubo formula based on retarded correlation functions.
}

\begin{document}
\maketitle

\section{Introduction}
\label{sec:Intro}
Spin in the non-relativistic limit, where the spin-orbit coupling is suppressed, becomes an approximately conserved quantity (good quantum number). 
This is the case for the heavy quarks in QCD (e.g.,~charm quark); the resulting heavy-quark symmetry shows up as the degeneracy in the spectrum of particular hadrons~\cite{Isgur:1989vq,Isgur:1990yhj,Isgur:1991wq,Neubert:1993mb,Manohar-Wise2000}. 
Another familiar realization is a certain class of condensed matter systems (e.g., magnetic materials), which has motivated recent development of the  hydrodynamic theory describing transport phenomena of spin, leading to a fruitful research area known as  {\it spintronics} (see Ref.~\cite{maekawa2017spin} for a review).
On the other hand, spin in {\it relativistic} systems is not a conserved quantity even approximately -- It is only a part of the total angular momentum, and is not conserved due to the spin-orbit coupling inherent in relativistic dynamics, whose corresponding transport theory has not been formulated until recently.

However, the recent experimental observation of spin polarization of hadrons in relativistic heavy-ion collisions~\cite{STAR:2017ckg,STAR:2019erd,ALICE:2019aid,STAR:2020xbm} strongly 
motivates the development of the theory describing spin transport in relativistic plasma, in particular, the quark-gluon plasma (QGP).
This motivation has led to several theoretical studies of relativistic hydrodynamics with spin polarization, based on the second law of thermodynamics~\cite{Hattori:2019lfp,Fukushima:2020ucl,Li:2020eon,She:2021lhe,Gallegos:2021bzp}, equilibrium partition functions~\cite{Gallegos:2021bzp}, quantum kinetic theory of relativistic fermions~\cite{Florkowski:2017ruc,Gao:2019znl,Hattori:2019ahi,Li:2019qkf,Yang:2020hri,Weickgenannt:2020aaf,Liu:2020flb,Bhadury:2020puc,Shi:2020htn,Peng:2021ago}, 
holographic approach for strongly-coupled plasma~\cite{Hashimoto:2013bna,Garbiso:2020puw,Gallegos:2020otk}, effective Lagrangian approach~\cite{Montenegro:2017rbu,Montenegro:2017lvf,Montenegro:2018bcf,Montenegro:2020paq}, and quantum statistical density operators~\cite{Becattini:2007nd,Becattini:2009wh,Becattini:2012pp,Becattini:2018duy,Hu:2021lnx} (see also Refs.~\cite{Florkowski:2018fap,Becattini:2020sww,Speranza:2020ilk} and references therein for a review).
These works have shed light on different aspects of relativistic hydrodynamics including spin degrees of freedom, often referred to as 
({\it relativistic}) {\it spin hydrodynamics}.

Spin angular momentum in the hydrodynamic regime of long time and distance scales is in general not a conserved quantity, due to exchanges with orbital angular momentum via spin-orbit coupling. 
However, the total angular momentum is conserved. As a result, an equilibrium state is characterized by a corresponding thermodynamic parameter~\cite{Landau:Statistical}
-- the thermodynamic conjugate to the total angular momentum, which we refer to as the {\it angular momentum ``chemical'' potential}. 
In the absence of torsion, this chemical potential
is equal to the thermal fluid vorticity \cite{Becattini:2007nd,Becattini:2009wh}, see Appendix \ref{sec:equilibrium}.
In a globally rotating (torsion-free) equilibrium, the angular momentum chemical potential, i.e. the thermal vorticity, is constant in space.  
Since there is only one angular momentum chemical potential for the conserved total angular momentum, both spin and orbital angular momenta in equilibrium  must be determined uniquely by that angular momentum chemical potential.  
That is, the spin polarization in equilibrium is fixed by the thermal vorticity, and their relation is one of the equilibrium thermodynamic properties.

On the other hand, since the system has a finite microscopic correlation length, the state of a local fluid element and its time evolution can only experience the local environment, and hence should be determined by the local thermal vorticity as well as other local thermodynamic parameters such as temperature and chemical potentials of conserved charges. This means that the fluid element in off-equilibrium relaxes to ``local'' equilibrium, with the thermal vorticity as one of the parameters characterizing local thermodynamic equilibrium.%
\footnote{
We note that the term ``local equilibrium'' has been used in literature in different contexts. Here, we use it to specify the local state where the strict hydrodynamic description can apply. In Refs.~\cite{Becattini:2012pp,Becattini:2018duy,Speranza:2020ilk}, however, the same term was used to refer to the state in which local entropy density is maximized with fixed local densities of conserved quantities, where the angular momentum ``chemical'' potential is not necessarily equal to the local thermal vorticity.}
In other words, spin polarization in the strict regime of hydrodynamics based on local equilibrium is not an independent variable, but is slaved to the conventional hydrodynamic variable of fluid velocity. Any deviation of spin polarization from its local equilibrium value is a non-hydrodynamic mode, which relaxes to zero with a finite relaxation rate that is determined by microscopic dynamics of the theory~\cite{Li:2019qkf,Kapusta:2019sad,Yang:2020hri, Weickgenannt:2020aaf,Weickgenannt:2021cuo}, similar to any other non-hydrodynamic modes. 
In a theory where the relaxation rate of spin polarization is much slower than other non-hydrodynamic modes, one can have two distinct hydrodynamic descriptions of spin polarization, depending on the time scale of interest compared to the relaxation time of spin polarization.

When the relevant time scale is much longer than the spin relaxation time, the spin is in local equilibrium with the thermal vorticity, and is not an independent degree of freedom. 
The system is described in a strict sense by hydrodynamics~\cite{Landau:Fluid}.  
The true novelty of spin hydrodynamics in this regime is two-fold~\cite{Li:2020eon}: 1) the spin, or equivalently the fluid vorticity, affects the local thermodynamic laws used in hydrodynamics, as a second order gradient correction to the first law of thermodynamics, 2) the energy-momentum tensor has an anti-symmetric part which is proportional to the rate of change of the spin tensor, i.e., the fluid vorticity. 
One can formulate spin hydrodynamics in this regime based on the gradient expansion of thermal vorticity. In leading order it gives the ideal limit of spin hydrodynamics with no production of entropy~\cite{Li:2020eon}, which is in accord with the existence of globally rotating equilibrium  with a constant, but non-zero thermal vorticity. Interestingly, this ideal spin hydrodynamics has been shown to be equivalent, by a pseudo-gauge transformation~\cite{Becattini:2012pp,Becattini:2018duy,Florkowski:2018fap,Speranza:2020ilk}, to conventional hydrodynamics based on the symmetric energy-momentum tensor with no spin degrees of freedom, with certain non-dissipative second order transport coefficients involving fluid vorticity~\cite{Li:2020eon}. The equivalence provides an important conceptual bridge between the two different formulations in the strict regime of hydrodynamics.

On the other hand, in the regime where  the time scale of interest is comparable to or shorter than the spin relaxation time (but is still longer than relaxation time of other non-hydrodynamic modes), one can include spin polarization as an additional independent dynamical mode~\cite{Hattori:2019lfp} in an extended hydrodynamic framework generally called Hydro+\cite{Stephanov:2017ghc}.
The essential feature of this extended framework is that spin polarization as a non-hydrodynamic mode relaxes to its local equilibrium value, due to spin-orbit coupling inherent in the microscopic relativistic theory~\cite{Hattori:2019lfp,Weickgenannt:2020aaf}.
A new transport coefficient appears in this regime, dictated by the second law of thermodynamics~\cite{Hattori:2019lfp,Fukushima:2020ucl,She:2021lhe,Gallegos:2021bzp}: the rotational viscosity, which determines the characteristic time scale for spin relaxation.
In the present paper, we focus on this regime of spin hydrodynamics.  

We address several important theoretical issues, founded upon the microscopic definition of the spin tensor in a background with torsion.
First, the definition of spin current has been unclear in some of the previous works. Although the canonical spin current of Dirac fermions  based on quantum field theory is totally anti-symmetric with respect to its three Lorentz indices, that was often not assumed in previous works.%
\footnote{One can perform a pseudo-gauge transformation to bring the spin current into a desired form, which however changes the definition of the spin current.}
Therefore, it is desirable to work with a microscopic definition of the spin-current operator from the underlying theory, possessing the appropriate anti-symmetric property of the spin-current.  
Second, the regime of applicability of relativistic spin hydrodynamics is rather subtle. A naive application of derivative expansion is problematic, and the proper expansion scheme needs to be clarified. 
Last but not least is the issue of how rapidly the spin density in a relativistic plasma relaxes to its equilibrium -- one of the most crucial information relevant to the ongoing heavy-ion experiments.
To answer this question, we need theoretical methods to evaluate the spin relaxation rate in the microscopic theory. One approach is the quantum kinetic theory with collisions~\cite{Li:2019qkf,Kapusta:2019sad,Yang:2020hri, Weickgenannt:2020aaf,Weickgenannt:2021cuo}. 
In this work, we follow an alternative path that does not rely on weak-coupling approximation and leads to  the Green-Kubo formula based on retarded correlation functions of spin observables, which is similar to those for other transport coefficients~\cite{Green,Kubo,Nakano}, but also with important differences due to non-conservation of spin density.%
\footnote{
Ref.~\cite{Hu:2021lnx} has recently derived a Green-Kubo formula based on the statistical operator approach, which has some overlap with our results.
} 
Our technique involves torsion which  makes spin connection an independent source for the spin current 
and facilitates derivation of Green-Kubo formulae.

Our approach is based on the combination of entropy-current analysis and linear response theory.
Our starting point is the Ward-Takahashi identity for the local Lorentz symmetry in the underlying quantum field theory of Dirac fermions, such as QED and QCD.
The first-order formalism in the background geometry with torsion naturally leads to a totally anti-symmetric spin current, which is not conserved due to spin-orbit coupling.
As a result of the total anti-symmetric nature of the spin current, our relativistic spin hydrodynamics contains three non-hydrodynamic modes, corresponding to a spin vector that is the generator of spatial rotations.

We then perform the entropy-current analysis in a general background with torsion, which introduces a new transport coefficient, the rotational viscosity $\eta_s$. 
The constitutive relations we find in a general torsionful background allow us to derive the Green-Kubo formula for the rotational viscosity, as well as other possible ways to compute it from the underlying quantum field theory.

Furthermore, the constitutive relations  tell us that the local equilibrium value of the angular momentum chemical potential receives additional contributions coming from the background torsion.
We study the linear response theory of our extended hydrodynamics with spin modes, and obtain the dispersion relations of dynamical modes.
The relaxation rate of the spin modes is shown to be given by $\Gamma_s = 2 \etas/\chi_s$, where $\chi_s$ is the spin susceptibility (see (\ref{spinsus}) for the definition).

\begin{figure}[t]
 \centering
 \includegraphics[width=0.7\linewidth]{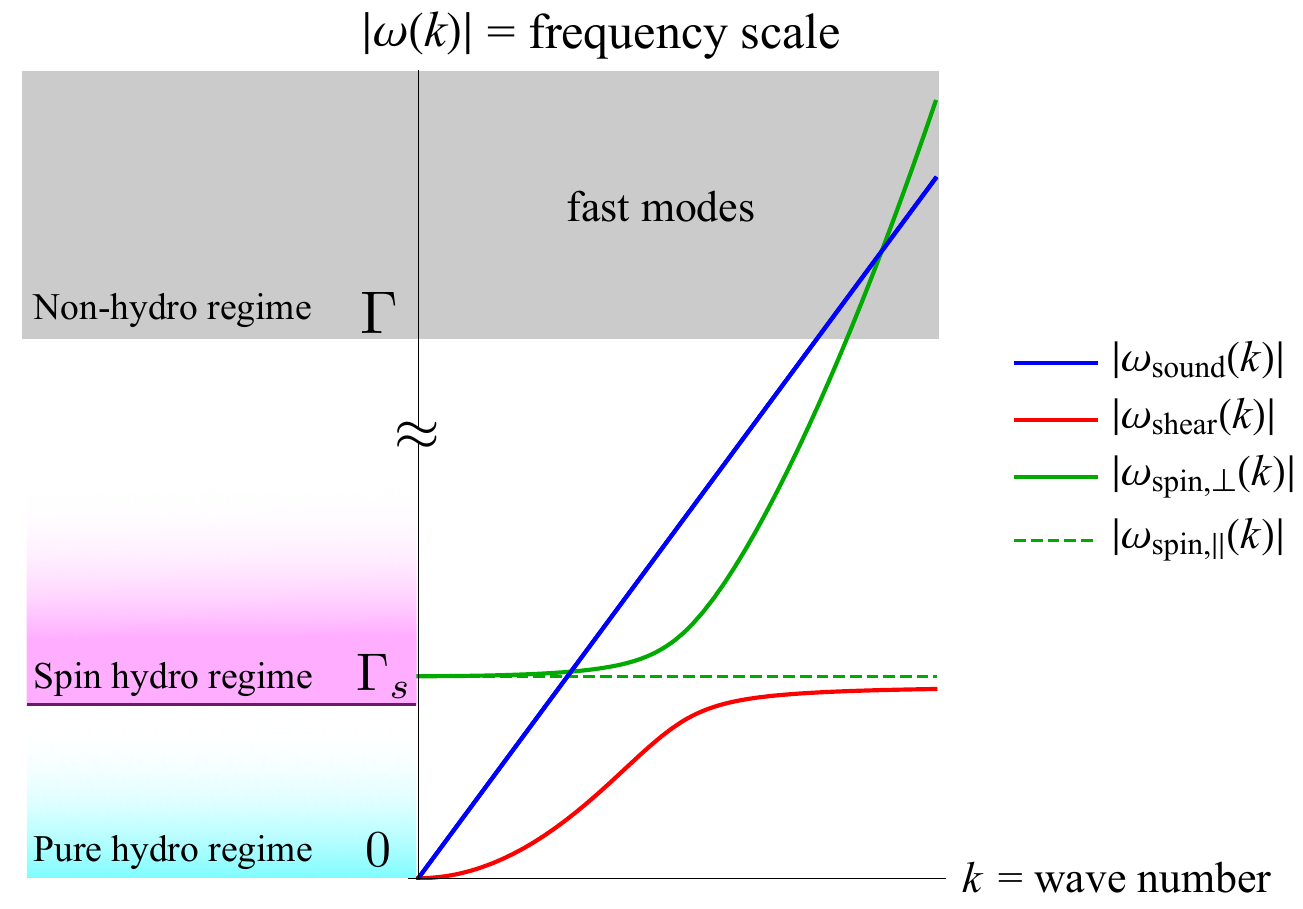}
 \caption{
A schematic picture of frequency scales as well as sound, shear, and spin density modes. The spin hydrodynamics discussed in this work extends the regime of validity of hydrodynamics to $\omega \gtrsim \Gamma_s$, where $\Gamma_s$ is the relaxation rate for spin density. For $\omega\ll \Gamma_s$, the spin is no longer an independent degree of freedom, and though spin hydrodynamics still applies, we also have pure hydrodynamics as an alternative and simpler description. The green and red lines show a mode mixing phenomenon between transverse shear and spin modes at the scale $\omega\gtrsim \Gamma_s$.
}
 \label{fig:energy-scale}
\end{figure}

We consider the frequency scale $\omega \gtrsim \Gamma_s $, where these spin modes are treated as independent dynamical variables in our relativistic spin hydrodynamics.
We also assume that $\omega\ll \Gamma$, where $\Gamma$ is the relaxation rate of other non-hydrodynamic modes. This window of scale, where our extended hydrodynamics with additional spin modes is justified, is possible only when $\Gamma_s$ is parametrically smaller than~$\Gamma$ (see Figure~\ref{fig:energy-scale}).
For example, in a weakly coupled quark-gluon plasma, when the mass $M$ of fermions is much larger than the temperature $T$, $\Gamma_s$ is expected to be smaller than $\Gamma$ by additional powers of $T/M$.

One of the interesting features we find in our linear response theory is the mixing between the transverse shear mode and the spin mode, that can be seen in Figure~\ref{fig:energy-scale}. This mode mixing arises because the transverse shear gradient $\partial_z u_x$ is a linear combination of shear tensor and fluid vorticity, the latter of which couples to spin modes.

The paper is organized as follows:
In section~\ref{sec:Spin}, we review the definition of energy-momentum tensor and spin current in quantum field theory of Dirac fermions, together with the associated Ward-Takahashi identity of local Lorentz symmetry.
In section~\ref{sec:entropy-current}, we derive the constitutive relations of relativistic spin hydrodynamics in a torsionful curved background, where spin density is treated as an independent dynamical variable.
In section~\ref{sec:linear}, we develop the linear response theory of our spin hydrodynamics, which gives the dispersion relation of dynamical modes, as well as the derivation of the Green-Kubo formula for the spin relaxation rate.
Section~\ref{sec:Summary} is devoted to summary and discussion.
In Appendices~\ref{sec:equilibrium} and \ref{sec:metric}, we present a discussion on the angular momentum in global equilibrium under rotation and
the relation between the currents in the vierbein-spin connection formalism and the metric-affine connection formalism, respectively.

\section{Spin current in the first-order formalism}
\label{sec:Spin}
In this section, we review the definition of the  energy-momentum tensor and the spin current from the viewpoint of the first-order (or Palatini) formalism for background spacetime~\cite{Palatini1919}.
We consider quantum field theory (QFT) in a torsionful (Einstein-Cartan) background geometry~\cite{Hehl:1976kj} and introduce currents and the Ward-Takahashi identities~\cite{Ward:1950xp,Takahashi:1957xn} associated with diffeomorphism, local Lorentz invariance and flavor symmetry.

\subsection{QFT in a torsionful background}
When the system enjoys a continuous global symmetry, it brings about a conserved current thanks to Noether's theorem~\cite{Noether1971}.
Nevertheless, the naive definition of the Noether current often contains conceptual problems, an example of which is the gauge non-invariance of the canonical energy-momentum tensor for gauge theories.
Although these problems may be resolved in an \textit{ad hoc} way by using, e.g., the Belinfante improvement~\cite{Belinfante1939,Belinfante1940,Rosenfeld1940}, it is not  clear whether the improved one is the best definition of current operators~(see, e.g., Refs.~\cite{Leader:2013jra,Wakamatsu:2014zza} and references therein for reviews on the definition of spin currents in gauge theories).

In order to avoid potential problems in the Noether current, 
one can instead utilize the definition of the \textit{gauge current} in the following systematic way: we introduce the background gauge field that couples to the global symmetry of the system, and by taking the variation of the gauged action with respect to the background gauge field, one immediately gets a well-behaved current operator.
A famous example is the symmetric and gauge-invariant energy-momentum tensor in general relativity, which is canonically defined by the variation of the matter action with respect to the metric.

In the same manner, we can define the spin current operator, although it is not a conserved current due to its possible exchanges with orbital angular momentum.
The key concept here is the spin connection, which enables us to promote the Lorentz invariance to a local one. 
Let us consider QFTs in a background of torsionful curved spacetime, whose action reads as
\begin{equation}
 \Scal 
 \big[\varphi; j\big]
  = 
  \int \diff^4 x e \Lcal  (\varphi,\partial_\mu \varphi;j),
  \label{eq:QCDaction}
\end{equation}
where we introduce a set of dynamical fields $\varphi$ and 
background fields $j$, respectively.
As an explicit example, we will consider QCD in this paper, but the whole discussion also applies to QED or any other theories including Dirac fermions.
We consider the following QCD Lagrangian density
\begin{equation}
 \Lcal_{\qcd}
  \equiv 
  - \frac{1}{2}\ptc{q}
  \Big( \gamma^{\ha} e_{\ha}^{~\mu} \ra{D}_\mu 
  - \la{D}_\mu e_{\ha}^{~\mu} \gamma^{\ha} \Big) q
  - \ptc{q} M q
  - \frac{1}{2} \tr 
  \left( g^{\mu\nu}g^{\alpha\beta}
   G_{\mu\alpha}G_{\nu\beta} \right),
  \label{eq:QCDLagrangian}
\end{equation}
where $q = (u,d,\cdots)^t$ denotes the quark fields
with $\ptc{q} \equiv \rmi q^\dag \gamma^0$, 
$M = \diag (m_u, m_d, \cdots)$ is their mass matrix, 
and $G_{\mu\nu} \equiv \partial_\mu a_\nu - \partial_\nu a_\mu 
- \rmi g [a_\mu,a_\nu]$ is the field strength tensor for the gluon field $a_\mu \equiv a_\mu^a t_a$, with $\mathrm{su}(3)_{\mathrm{c}}$ generators $t_a$, and the QCD coupling constant $g$.
The dynamical fields are 
quarks and gluons: $\varphi = \{q,a_\mu\}$, and 
corresponding to global symmetries of QCD,
a set of background fields is given by 
$j = \{e_\mu^{~\ha},\omega_\mu^{~\ha\hb},A_\mu\}$, where $e_\mu^{~\ha}~(e_{\ha}^{~\mu})$ and $\omega_\mu^{~\ha\hb} =-\omega_\mu^{~\hb\ha}$ denote the (inverse) vierbein and spin connection with $e \equiv \det e_\mu^{~\ha}$, respectively, and $A_\mu$ is a flavour gauge field, coupled to, e.g., the $\U(1)$ baryon symmetry (generalization to non-Abelian flavour symmetry is straightforward). 
We note that the anti-symmetry of $\omega_\mu^{~\hat a \hat b}$ implies metric compatibility, $\nabla_\alpha g_{\mu\nu}=0$. 
We also introduced the covariant derivative of the quark field as 
\begin{equation}
 \begin{split}
   \ra{D}_\mu q &= 
   \partial_\mu q - \rmi g a_\mu q 
   - \frac{\rmi}{2} \omega_\mu^{~\ha\hb} \Sigma_{\ha\hb} q
   - \frac{\rmi}{3} A_\mu q ,
   \\
  \bar{q} \la{D}_\mu &= 
   \partial_\mu \bar{q} + \rmi g \bar{q} a_\mu 
   + \frac{\rmi}{2} \omega_\mu^{~\ha\hb} \bar{q} \Sigma_{\ha\hb} 
   + \frac{\rmi}{3} A_\mu q ,
 \end{split}
  \label{eq:CovDerFermion}
\end{equation}
where $\gamma^{\ha}$ are the Dirac gamma matrices, and 
$\Sigma_{\ha\hb} \equiv \rmi [\gamma_{\ha},\gamma_{\hb}]/4$ are the generators of the  Lorentz group in the spinor representation.
In this paper, we use Greek (hatted Latin) letters for coordinate (local Lorentz) indices.
The Minkowski metric convention is chosen to be the mostly plus one: $\eta_{\ha\hb} = \diag (-1,+1,+1,+1)$.

It is worth emphasizing that the fundamental variables describing the background geometry are 
the vierbein and the spin connection (or contorsion which will be introduced shortly). 
We consider a torsionful background, where the vierbein and the spin connection are independent background fields.
One can then define the space-time metric $g_{\mu\nu}$ and the affine (curved space-time) connection $\Gamma^\rho_{~\mu\nu}$ through
\begin{equation}
 g_{\mu\nu} = e_\mu^{~\ha} e_\nu^{~\hb} \eta_{\ha\hb} 
  \quad \mathrm{and} \quad
  D_\mu e_\nu^{~\ha} =
  \partial_\mu e_\nu^{~\ha} - \Gamma^\rho_{~\mu\nu} e_\rho^{~\ha} 
  + \omega_{\mu~\hb}^{~\ha} e_\nu^{~\hb} = 0,
  \label{eq:g-e}
\end{equation}
where the second equation, or the so-called tetrad postulate, defines the affine connection from the given vierbein and spin connection.
The anti-symmetric part of the tetrad postulate gives the first Cartan equation, which defines the torsion $T^{\rho}_{~\mu\nu} \equiv \Gamma^\rho_{~\mu\nu} - \Gamma^\rho_{~\nu\mu}$ as follows:
\begin{equation}
 T^{\ha} = \diff e^{\ha} + \omega^{\ha}_{~\hb} e^{\hb} 
  \eqq 
  T^{\ha}_{~\mu\nu} 
  = \partial_{\mu} e_\nu^{~\ha} - \partial_{\nu} e_\mu^{~\ha} 
  + \omega_{\mu~\hb}^{~\ha} e_\nu^{~\hb} - \omega_{\nu~\hb}^{~\ha} e_\mu^{~\hb}.
\end{equation}

The introduction of the torsion is reasonable because it promotes the spin connection to an independent background field.
This becomes manifest by solving the tetrad postulate for the spin connection as
\begin{equation}
 \begin{split}
  \omega_\mu^{~\ha\hb} &= 
  \comega_\mu^{~\ha\hb} (e) + K_\mu^{~\ha\hb},
  \label{eq:SpinC}
 \end{split}
\end{equation}
where we introduced the torsion-free spin connection 
$\comega_\mu^{~\ha\hb} (e)$ and the contorsion $K_\mu^{~\ha\hb}$ as follows~\cite{Hehl:1976kj}:
 \begin{align}
 \comega_\mu^{~\ha\hb}
 &\equiv \frac{1}{2}
  e^{\ha\nu} e^{\hb\rho} 
  (- C_{\mu\nu\rho} + C_{\nu\rho\mu} - C_{\rho\nu\mu} ),
 \\
 K_\mu^{~\ha\hb}
 &\equiv 
  \frac{1}{2}
  e^{\ha\nu} e^{\hb\rho} 
  ( T_{\mu\nu\rho} - T_{\nu\rho\mu} + T_{\rho\nu\mu}) ,
\end{align}
where $C_{\mu\nu\rho}$ is the Ricci rotation coefficient given by 
\begin{equation}
 C_{\mu\nu\rho} \equiv e_\mu^{~\hc} 
  (\partial_\nu e_{\rho \hc} - \partial_{\rho} e_{\nu \hc}).
\end{equation}
Note that the torsion-free part $\comega_\mu^{~\ha\hb} (e)$ is fixed by the vierbein.
Thus, introducing the spin connection as an independent field is equivalent to putting the system into a background with torsion.
One observes that the quark couples to the background torsion through the spin connection.

It is worthwhile to emphasize that the gluon (or photon in the case of QED) is assumed to have no coupling to the torsion, in contrast to the quark field.
This is a consequence of imposing $\SU(3)_{\mathrm{c}}$ gauge invariance of the action~\cite{Hehl:1976kj}.
Note that we define the gluon field strength tensor with a simple partial derivative, not with the covariant derivative.
At first glance, this choice does not seem compatible with diffeomorphism invariance, but it indeed is.
To see this, notice that both covariant derivatives with or without  torsion are consistent with diffeomorphism invariance, since their difference is proportional to the torsion \textit{tensor}.
However, one sees that the use of the covariant derivative with torsion for the field strength tensor spoils $\SU(3)_{\mathrm{c}}$ gauge invariance 
due to the last term appearing in 
\begin{equation}
 \nabla_\mu a_\nu - \nabla_\nu a_\mu 
  - \rmi g [a_\mu,a_\nu]
  = \partial_\mu a_\nu - \partial_\nu a_\mu 
  - \rmi g [a_\mu,a_\nu]
  - T^\rho_{~\mu\nu} a_\rho .
\end{equation}
Throughout the paper, we use $\nabla_\mu$ as the covariant derivative including only the affine connection, and $D_\mu$ as the one including both the affine and spin connections.
It should also be pointed out that we have the symmetrized derivative in the fermion kinetic term in Eq.~\eqref{eq:QCDLagrangian}. 
This symmetrized form ensures that the Lagrangian is real-valued, 
and as a consequence, the resulting spin current is a Hermitian operator 
as it should be.
These choices are crucial for the proper definition of the spin current, or equivalently, for our decomposition of the total angular momentum into spin and orbital angular momenta.

\subsection{The Ward-Takahashi identities}
\label{sec:Conservation}
The systems enjoying the Poincar\'e invariance and some flavour symmetries are equipped with conserved charges, given by the energy, momentum, total angular momentum, and flavour charges.
Since we would like to discuss the dynamics of spin polarization, we need to identify the proper definition of the spin current,
in order to decompose the total angular momentum into the orbital and the spin parts.
The decomposition of the angular momentum tensor in quantum gauge theories is still a controversial subject (see, e.g., Refs.~\cite{Leader:2013jra,Wakamatsu:2014zza} and references therein for a review).%
\footnote{
In our view, any decomposition should be acceptable as long as it respects $\SU(3)_{\mathrm{c}}$ gauge invariance.
The real issue is what component we measure in a given experiment.
}
Based on the background vierbein and spin connection, we here introduce one way to define a gauge-invariant spin current,
which eventually is shown to involve only the fermion sector.

First of all, note that the introduction of background vierbein and spin connection promotes the Poincar\'e symmetry to local symmetries --- diffeomorphism and local Lorentz invariance.
As a consequence of these local symmetries, one can derive in the usual manner the covariant (non-)conservation laws as the Ward-Takahashi identities.
The spatial components of the associated spin current, i.e. $\Sigma^{ijk}$, turn out to be completely determined by the axial charge density $J_{5}^0$.   
Also, it gives us a useful prescription to define the spin current through the variation of the QCD action \eqref{eq:QCDaction}-\eqref{eq:QCDLagrangian} with respect to the spin connection.
One can regard this procedure as a familiar way to define the gauge current.
We thus define the ``canonical''  energy-momentum tensor $\Theta^\mu_{~\ha}$ and the spin current $\Sigma^\mu_{~\ha\hb}$, 
as well as the flavour current $J^\mu$, 
by taking the variation with respect to $e_\mu^{~\ha}$ and $\omega_{\mu}^{~\ha\hb}$, 
as well as $A_\mu$, 
respectively:
\begin{equation}
 \Theta^\mu_{~\ha} (x) \equiv 
  \frac{1}{e (x)} 
  \left. \frac{\delta \Scal_{\qcd}}{\delta e_\mu^{~\ha} (x)} \right|_{\omega,\, A}, 
  \quad
  \Sigma^{\mu}_{~\ha\hb} (x)
  \equiv 
  - \frac{2}{e(x)} 
   \left. 
    \frac{\delta \Scal_{\qcd}}{\delta \omega_{\mu}^{~\ha\hb}(x)}
   \right|_{e,\, A}, 
  \quad
  J^\mu (x) \equiv 
  \frac{1}{e (x)} 
  \left .
  \frac{\delta \Scal_{\qcd}}{\delta A_\mu (x)} \right|_{e,\,\omega},
\label{eq:Currents}
\end{equation}
where subscripts $\omega$ and $e$ indicate that they are fixed when we compute the functional derivative.

After a straightforward calculation, we find that the QCD action \eqref{eq:QCDaction} gives us the following expression for our currents
\begin{equation}
 \begin{split}
  \Theta^\mu_{~\ha}
  &= \frac{1}{2} 
  \bar{q} \big( \gamma^\mu \ra{D}_{\ha} - \la{D}_{\ha} \gamma^\mu \big) q
  + 2 \tr \left( G^{\mu\rho} G_{\ha\rho} \right) 
  + \Lcal_{\qcd} e^{~\mu}_{\ha},
  \\
  \Sigma^\mu_{~\ha\hb}
  &= - \frac{\rmi}{2} \ptc{q} e^\mu_{~\hc}
  \{ \gamma^{\hc}, \Sigma_{\ha\hb} \} q,
  \\
  J^\mu
  &= \frac{\rmi}{3} \ptc{q} e^\mu_{~\ha} 
  \gamma^{\ha} q,
 \end{split}
 \label{eq:Current-expression}
\end{equation}
where $\{A,B\} \equiv  AB + BA$ is the anti-commutator.
One finds that only the local (or on-site) quark field contributes to the spin current, so that our spin current is manifestly $\SU(3)_{\mathrm{c}}$ gauge-invariant.
Moreover, thanks to the anti-commutator,  
$\Sigma^\mu_{~\ha\hb}$ are Hermitian gauge-invariant operators, as all physical observables should be.
In other words, the spin current would not be a Hermitian operator, if the fermion kinetic term in the Lagrangian was not symmetrized. 
The absence of gluon contributions to the spin current results from the absence of the torsion in the gluon field strength tensor due to gauge invariance.
In that sense, our requirement of $\SU(3)_{\mathrm{c}}$ gauge invariance in torsionful geometry fixes our choice of the decomposition of the angular momentum.

There is another useful expression for the spin current, thanks to a gamma matrix identity
$\{\gamma^\mu, \Sigma^{\nu\rho} \} =\vep^{\mu\nu\rho\sigma} 
\gamma_\sigma \gamma_5$ with $\gamma_5 \equiv - \rmi \gamma^{\hzero} \gamma^{\hone} \gamma^{\htwo} \gamma^{\hthree}$, where the normalized totally anti-symmetric tensor is defined as
$\vep^{\mu\nu\rho\sigma} \equiv \epsilon^{\mu\nu\rho\sigma}/e$ with $\epsilon^{0123} = +1$. 
This formula tells us that the spin current is totally anti-symmetric with respect to its three indices.
One can then easily show that it is equal to the Hodge dual of the axial $\U(1)_A$ current $J_5^{\ha}$ as 
\begin{equation}
 \Sigma^{\mu}_{~\ha\hb} 
  = - \frac{1}{2} \vep^{\mu}_{~\ha\hb\hc}
  J_{5}^{\hc}
  \with 
  J_5^{\ha} \equiv 
  \rmi \ptc{q} \gamma^{\ha} \gamma_5 q.
  \label{eq:chiral}
\end{equation}
Therefore, the number of independent components of the spin current operator is only four, which are in one-to-one correspondence to the $\U(1)_A$ current $J_5^\mu$.
This means that some components of the spin density, e.g. $\Sigma^{00i}$, are identically zero, and the other components such as $\Sigma^{0ij}$ are equivalent to 
the spatial components of the axial current $J_5^i$.
The remaining component of $\Sigma^{ijk}$ is completely determined by the axial charge density $J_{5}^0$.

Let us now derive the Ward-Takahashi identities that follow from local symmetries.
The vital point is that the QCD action~\eqref{eq:QCDaction} remains invariant, i.e. $\delta_\chi \Scal_{\qcd} = 0 $, under the following infinitesimal transformations:
\begin{equation}
 \begin{cases}
  \delta_\chi e_\mu^{~\ha}
  = \xi^\nu \partial_\nu e_\mu^{~\ha} + e_\nu^{~\ha} \partial_\mu \xi^\nu 
  - \alpha^{\ha}_{~\hb} e_\mu^{~\hb} ,
  \\  
  \delta_\chi \omega_{\mu~b}^{~\ha}
  = \xi^\nu \partial_\nu \omega_{\mu~\hb}^{~\ha} 
  + \omega_{\nu~\hb}^{~\ha} \partial_\mu \xi^\nu 
  + \partial_\mu \alpha^{\ha}_{~\hb}
  - \alpha^{\ha}_{~\hc} \omega_{\mu~\hb}^{~\hc} 
  + \alpha^{\hc}_{~\hb} \omega_{\mu~\hc}^{~\ha},
  \vspace{3pt} \\
 \delta_\chi A_\mu 
  = \xi^\nu \partial_\nu A_\mu + A_\nu \partial_\mu \xi^\nu
  + \partial_\mu \theta,
  \\
  \delta_\chi q = \xi^\nu \partial_\nu q 
  + \dfrac{\rmi}{2} \alpha^{\ha\hb} \Sigma_{\ha\hb} q
  + \dfrac{\rmi}{3} \theta q ,  \vspace{3pt} \\
  \delta_\chi \bar{q} = \xi^\nu \partial_\nu \ptc{q}
  - \dfrac{\rmi}{2} \alpha^{\ha\hb} \bar{q} \Sigma_{\ha\hb} 
  - \dfrac{\rmi}{3} \theta q ,
  \\
  \delta_\chi a_\mu 
  = \xi^\nu \partial_\nu a_\mu + a_\nu \partial_\mu \xi^\nu.
 \end{cases}
\end{equation}
We introduced a set of local infinitesimal parameters, $\chi \equiv \{\xi^\mu, \alpha^{\ha}_{~\hb}, \theta\}$, which generates the general coordinate and local Lorentz transformations -- four-vector $\xi^\mu$ for diffeomorphism, anti-symmetric tensor $\alpha_{\ha\hb} = - \alpha_{\hb\ha}$ for local Lorentz symmetry, and 
$\theta$ for the $\U(1)$ symmetry.
On the other hand, a direct computation of the induced variation of the QCD action leads to another expression for $\delta_\chi \Scal_{\qcd}$ 
with the arbitrary parameters $\chi$. 
For $\delta_\chi \Scal_{\qcd} = 0$ to hold for arbitrary $\chi$, 
their coefficients need to vanish, which gives us the Ward-Takahashi identities for the currents.
Since the computation is a little complicated, we first consider the identities associated with the local Lorentz and $\U(1)$ symmetries, and then move on to that for diffeomorphism.

The variation of the action under the local Lorentz transformation is given by
\begin{align}
  \delta_\alpha \Scal_{\qcd}
  &= \int \diff^4 x
  \left[
  \frac{\delta \Scal_{\qcd}}{\delta e_\mu^{~\ha} } 
  \delta_\alpha e_\mu^{~\ha} 
  + \frac{\delta \Scal_{\qcd}}{\delta \omega_{\mu~\hb}^{~\ha} } 
  \delta_\alpha \omega_{\mu~\hb}^{~\ha} 
  + \frac{\delta \Scal_{\qcd}}{\delta q} 
  \delta_\alpha q
  + \frac{\delta \Scal_{\qcd}}{\delta \bar{q}} 
  \delta_\alpha \bar{q}
  + \frac{\delta \Scal_{\qcd}}{\delta a_\mu} 
  \delta_\alpha a_\mu
  \right]
 \nonumber \\
  &= \frac{1}{2} \int \diff^4 x e 
  \alpha^{\ha\hb}
  \bigg[ (\Theta_{\ha\hb} - \Theta_{\hb\ha} ) 
  +  ( D_\mu - \Gcal_\mu ) \Sigma^{\mu}_{~\ha\hb} 
  \bigg] ,
 \label{eq:LL-variation}
\end{align}
where we used the equation of motion for dynamical variables 
($ \delta \Scal_{\qcd}/\delta q 
 = \delta \Scal_{\qcd}/\delta \bar{q}
 = \delta \Scal_{\qcd}/\delta a_\mu = 0 $). 
We here introduced a torsional contribution 
$\Gcal_\mu \equiv T^\nu_{~\nu\mu}$, which results from 
the integration by parts (all surface terms are neglected) in the presence of the torsion as 
\begin{equation}
 \partial_\mu e = e \Gamma^\nu_{~\mu\nu} 
  = e ( \Gamma^\nu_{~\nu\mu} - T^\nu_{~\nu\mu})
  = e ( \Gamma^\nu_{~\nu\mu} - \Gcal_{\mu}).
\end{equation}
The local Lorentz invariance of the action tells us $\delta_\alpha \Scal_{\qcd} = 0 $ holds for arbitrary position-dependent $\alpha^{\ha\hb}$.
As a result, we obtain the following identity:
\begin{equation}
\label{eq:spinCurrentConservation}
 (D_\mu - \Gcal_\mu) \Sigma^{\mu}_{~\ha\hb}
  = - (\Theta_{\ha\hb} - \Theta_{\hb\ha}).
\end{equation}
Therefore, based on Eq.~\eqref{eq:spinCurrentConservation}, the covariant divergence of our spin current is tied to the anti-symmetric part of our energy-momentum tensor.
We regard this Ward-Takahashi identity as the equation of motion for the spin density.
However, we also note that not all the components of 
Eq.~\eqref{eq:spinCurrentConservation} give
equations of motion for the spin density. 
This is because our spin current is totally anti-symmetric, and thus, we only have three spin components as dynamical variables.
As a result, the remaining three equations are shown to be constraints for a part of the energy-momentum tensor $\Theta^\mu_{~\ha}$.
We will use these constraints when we derive the 
constitutive relations with the help of the second law of local thermodynamics.
A similar computation leads to the following identity 
attached to the $\U(1)$ symmetry:
\begin{equation}
 (\nabla_\mu - \Gcal_\mu ) J^\mu = 0 
 \label{eq:WT-Baryon},  
\end{equation}
which gives a conservation law for the flavour current.

Being equipped with the above identities, we further evaluate the variation of the action induced by the general coordinate transformation as
\begin{equation}
 \begin{split}
  \delta_\xi \Scal_{\qcd}
  &=
  - \int \diff^4 x e \xi^\nu \left[
  ( \nabla_\mu - \Gcal_\mu ) \Theta^\mu_{~\nu}  
  + \Theta^\mu_{~\rho} T^{\rho}_{~\mu\nu}
  - \frac{1}{2} \Sigma^{\mu~\hb}_{~\ha} \Rcal^{\ha}_{~\hb\mu\nu}
  - F_{\nu\mu} J^\mu
  \right] ,
 \end{split}
 \label{eq:Diff-variation}
\end{equation}
where we again used the equation of motion and performed the integration by parts.
Here, we introduced the field strength tensor $\Rcal^a_{~b\mu\nu}$ 
attached to the spin connection, and $F_{\mu\nu}$ as 
\begin{equation}
 \begin{split}
   \Rcal^{\ha}_{~\hb\mu\nu} 
   &\equiv 
  \partial_\mu \omega_{\nu~\hb}^{~\ha}  - \partial_\nu \omega_{\mu~\hb}^{~\ha}  
  + \omega_{\mu~\hc}^{~\ha} \omega_{\nu~\hb}^{~\hc}
  - \omega_{\nu~\hc}^{~\ha} \omega_{\mu~\hb}^{~\hc} ,
  \\
  F_{\mu\nu} 
  &\equiv 
  \partial_\mu A_\nu - \partial_\nu A_\mu.
 \end{split}
\end{equation}
For $\delta_\xi \Scal_{\qcd} = 0 $ to hold for arbitrary $\xi^\nu$, 
we obtain the identity 
\begin{equation}
 ( \nabla_\mu - \Gcal_\mu ) \Theta^\mu_{~\nu}
  = - \Theta^\mu_{~\rho} T^\rho_{~\mu\nu}
 + \frac{1}{2} \Sigma^{\mu~\hb}_{~\ha} \Rcal^{\ha}_{~\hb\mu\nu}
  +F_{\nu\mu} J^\mu 
  .
  \label{eq:Diffeo}
\end{equation}
Therefore, our energy-momentum tensor has source terms coming from the spacetime torsion, curvature, and external $\U(1)$ gauge field.

In summary, we have obtained the following Ward-Takahashi identities:
\begin{equation}
 \begin{split}
  ( D_\mu - \Gcal_\mu ) \Theta^\mu_{~\ha}
  &= - \Theta^\mu_{~\hb} T^{\hb}_{~\mu\ha}
   + \frac{1}{2} \Sigma^{\mu~\hc}_{~\hb} \Rcal^{\hb}_{~\hc\mu\ha} 
   + F_{\ha\mu} J^\mu, 
   \\
  (D_\mu - \Gcal_\mu) \Sigma^{\mu}_{~\ha\hb}
  &= - ( \Theta_{\ha\hb} - \Theta_{\hb\ha} ),     
  \\
  (\nabla_\mu - \Gcal_\mu) J^\mu 
  &= 0 ,
 \end{split} 
 \label{eq:WT}
\end{equation}
which result from diffeomorphism, local Lorentz invariance and $\U(1)$ symmetry of the QCD action \eqref{eq:QCDaction}-\eqref{eq:QCDLagrangian}.
When we are interested in QCD in the flat spacetime with vanishing $\U(1)$ gauge field, we turn off the background fields $e_\mu^{~\ha}$, $\omega_\mu^{~\ha\hb}$, and $A_\mu$, which reduces Eqs.~\eqref{eq:WT} to 
\begin{equation}
 \partial_\mu \Theta^\mu_{~\nu} = 0 
   ,\quad
   \partial_\mu \Sigma^{\mu}_{~\ha\hb} 
   = - (\Theta_{\ha\hb} - \Theta_{\hb\ha}),
   \quad \mathrm{and} \quad 
   \partial_\mu J^\mu = 0.
   \label{eq:flat}
\end{equation}
These equations give the conservation laws for the 
canonical-like energy-momentum density, total angular momentum, and flavour charge (the first two were used in Ref.~\cite{Hattori:2019lfp}).
We note that the spin current $\Sigma^{\mu}_{~\ha\hb}$ is totally anti-symmetric with respect to its indices owing to the gamma matrix structure of Eq.~\eqref{eq:Current-expression} [recall Eq.~\eqref{eq:chiral}], which is not assumed in Ref.~\cite{Hattori:2019lfp}. 
As a result of the anti-symmetric property, the number of independent degrees of freedom in the spin current is not six but three in the present work.

One remark here is that we defined the energy-momentum tensor $\Theta^\mu_{~\ha}$ by considering the 
variation with the fixed spin connection $\omega_\mu^{~\ha\hb}$ instead of the fixed contorsion $K_\mu^{~\ha\hb}$.
As a consequence, one finds that $\Theta^\mu_{~\ha}$ corresponds to the canonical-like energy-momentum tensor. 
If we introduce the energy-momentum tensor $T^\mu_{~\ha}$ with the fixed contorsion $K_\mu^{~\ha\hb}$, we have 
\begin{equation}
 \begin{split}
  T^\mu_{~\ha} 
  &\equiv 
  \frac{1}{e(x)}
  \left. \frac{\delta \Scal_{\qcd}}{\delta e_\mu^{~\ha} (x)} \right|_{K}
  = \Theta^\mu_{~\ha} 
  + \frac{1}{2} (D_\rho - \Gcal_\rho) 
   \Sigma^{\mu~\rho}_{~\ha},
   \end{split}
 \label{eq:EM-T}
\end{equation}
where we used the totally anti-symmetric property of the spin current to obtain the rightmost side.
We will see that this energy-momentum tensor becomes useful when performing the linear-mode analysis in Sec.~\ref{sec:linear}.

\section{Derivation of spin hydrodynamics}
\label{sec:entropy-current}

In this section, we provide a derivation of hydrodynamics with spin polarization~\cite{Hattori:2019lfp,Fukushima:2020ucl,She:2021lhe,Gallegos:2021bzp} 
(see also \cite{Gallegos:2020otk}). 
Based on the Ward-Takahashi identities and the second law of local thermodynamics~\cite{Landau:Fluid}, we specify a first-order constitutive relation of hydrodynamics with the totally anti-symmetric spin current.
In addition to two (shear and bulk) viscosities 
and a flavour charge conductivity, we find 
one kinetic coefficient, called \textit{rotational viscosity}, appearing in the anti-symmetric part of the energy-momentum tensor.

\subsection{Setup}
Let us first introduce the dynamical variables. 
We employ the Landau-Lifshitz frame with respect to the symmetric part of the energy-momentum tensor in defining the fluid four-velocity $u^\mu$.
Therefore, we introduce $u^\mu$ and the associated energy density $\epsilon$ in the rest frame by 
\begin{equation}
 \Theta^\mu_{~\ha} \big|_{(s)} u^{\ha} 
 = - \epsilon u^{\mu}
  \with
  \eta_{\ha\hb} u^{\ha} u^{\hb} 
  = g_{\mu\nu} u^{\mu} u^{\nu} = - 1,
\end{equation}
where we denote the symmetric part of the energy-momentum tensor as 
$\Theta^\mu_{~\ha} \big|_{(s)} 
\equiv (\Theta^\mu_{~\ha} + \Theta^{~\mu}_{\ha})/2$.
Owing to the normalization $u_\mu u^\mu=-1$, we can parameterize the four-velocity as 
$u^\mu = \gamma (\bv) (1,\bv)^t$ with $\gamma (\bv) = 1/\sqrt{1 - \bv^2} $.
We next introduce the spin density $\sigma_{\ha\hb} = - \sigma_{\hb\ha}$ and the flavour charge density $n$ by 
\begin{equation}
 \sigma_{\ha\hb} = - u_\mu \Sigma^{\mu}_{~\ha\hb}
 \quad \mathrm{and} \quad
 n = - u_\mu J^\mu .
  \label{eq:spin-density}
\end{equation}
As given in Eq.~\eqref{eq:Current-expression}, we microscopically know the spin current for QFTs with Dirac fermions is totally anti-symmetric so that the spin density for such QFTs satisfies $\sigma_{\ha\hb} u^{\ha} = 0$.
This means that we only have three independent components of spin density attached to the spatial rotation symmetry, while the other three associated with the boost symmetry are absent.
Thus, our dynamical variables are three spin components in addition to four energy-momentum densities and one flavour charge density.
As for the spin density, we also use the dual variable $\sigma^{\ha}$ defined by 
\begin{equation}
 \sigma^{\ha}  
 \equiv \frac{1}{2}
 \vep^{\ha\hb\hc\hd} u_{\hb} \sigma_{\hc\hd}
 \eqq
  \sigma_{\ha\hb} = 
  -
  \vep_{\ha\hb\hc\hd} u^{\hc} \sigma^{\hd}
  \quad \mathrm{satisfying} \quad
  \sigma^{\ha} u_{\ha} = 0 = \sigma_{\ha\hb} u^{\ha}.
  \label{eq:dual-sigma}
\end{equation}
In short, we identify eight dynamical variables as
\begin{equation}
 \epsilon,~n,~u^{\ha},~\sigma^{\ha}~  (\mathrm{or}~ \sigma_{\ha\hb})
  \quad \mathrm{satisfying} \quad
  \eta_{\ha\hb} u^{\ha} u^{\hb} = -1,  \quad
  \sigma^{\ha} u_{\ha} = 0 = \sigma_{\ha\hb} u^{\ha} .
  \label{eq:dynamical-variables}
\end{equation}
We then introduce thermodynamic variables conjugate to the energy, spin, and flavour charge densities.
For that purpose, we introduce a generalized entropy density
$s(\epsilon,n,\sigma_{\ha\hb})$, or equivalently $s(\epsilon,n,\sigma^{\ha})$, and introduce the inverse temperature $\beta$, flavour chemical potential $\mu$,
and spin chemical potential $\mu^{\ha\hb}$ (or its dual $\mu_{\ha}$) by
\begin{equation}
 \beta \equiv \frac{\partial s}{\partial \epsilon} 
 , \quad
   \beta \mu \equiv
  -\frac{\partial s}{\partial n} ,
  \quad
  \beta \mu^{\ha\hb} 
   \equiv - 2 \frac{\partial s}{\partial \sigma_{\ha\hb}} \quad \mathrm{and} \quad 
  \left( 
   \mathrm{or} \quad 
   \beta \mu_{\ha} \equiv - \frac{\partial s}{\partial \sigma^{\ha}} 
  \right).
\end{equation}
In other words, we require the generalized first law of local thermodynamics in the rest frame including spin density: 
\begin{equation}
 T \diff s 
  = \diff \epsilon -\mu \diff n
  - \frac{1}{2} \mu^{\ha\hb} \diff \sigma_{\ha\hb}
  (=  \diff \epsilon -\mu \diff n- \mu_{\ha} \diff \sigma^{\ha}). 
\end{equation}
Equation \eqref{eq:dual-sigma} forces two spin chemical potentials to be related with each other by 
\begin{equation}
 \mu_{\ha} = 
 \frac{1}{2} \vep_{\ha\hb\hc\hd} u^{\hb} \mu^{\hc\hd}
 \eqq
 \mu^{\ha\hb} = - \vep^{\ha\hb\hc\hd} u_{\hc} \mu_{\hd}  
 . 
\end{equation}
Note that the spin chemical potential satisfies 
$\mu_{\ha} u^{\ha} = 0 = \mu^{\ha\hb} u_{\hb}$.

Our purpose is to express 
$\Theta^\mu_{~\ha}$, $\Sigma^\mu_{~\ha\hb}$, and $J^\mu$ by using the eight dynamical variables 
listed in Eq.~\eqref{eq:dynamical-variables}.
We then rewrite these currents by performing the tensor decomposition as follows:
\begin{equation}
 \begin{split}
  \Theta^\mu_{~\ha}
  &= \epsilon u^\mu u_{\ha} + p \Delta^{\mu}_{\ha} 
  + u^\mu \delta q_{\ha} - \delta q^\mu u_{\ha}
  + \delta \Theta^\mu_{~\ha}  ,
  \\
  \Sigma^\mu_{~\ha\hb}
  &= \vep^{\mu}_{~\ha\hb\hc} 
   ( \sigma^{\hc} + \delta \sigma u^{\hc}),
   \\
  J^\mu &= n u^\mu + \delta J^\mu,
 \end{split}
 \label{eq:general-constitutive}
\end{equation}
where we defined a pressure $p$ with a projection matrix $\Delta^\mu_{\ha}$:
\begin{equation}
 \Delta^\mu_{\ha} \equiv e^{~\mu}_{\ha} + u^\mu u_{\ha}
  \quad \mathrm{satisfying} \quad 
  \Delta^\mu_{\ha} u^{\ha} = 0 = \Delta^\mu_{\ha} u_{\mu}.
\end{equation}
Note that the Landau-Lifshitz definition of the fluid velocity restricts 
$\delta \Theta^\mu_{~\ha}$ and $\delta q_{\ha}$
to satisfy $\delta \Theta^\mu_{~\ha} u^{\ha} = 0 = 
u_\mu \delta \Theta^\mu_{~\ha}$ and 
$\delta q_{\ha} u^{\ha} = 0 $, 
and the definition of $n$ leads to $u_\mu \delta J^\mu = 0$. 
Equation \eqref{eq:general-constitutive} just gives one 
possible parameterization of $\Theta^\mu_{~\ha}$, $\Sigma^\mu_{~\ha\hb}$ and $J^\mu$, and thus, 
$\delta \Theta^\mu_{~\ha}$, $\delta q_{\ha}$, $\delta \sigma$, and $\delta J^\mu$ could contain nondissipative terms.
We will, however, show that such a nondissipative term is absent in the constitutive relation for 
$\delta \Theta^\mu_{~\ha}$, $\delta q_{\ha}$ and $\delta \sigma$ in the first-order spin hydrodynamics.

It is worth emphasizing that the frequency scale of our main interest is in the spin hydrodynamic regime, at which the spin density shows its intrinsic dynamics as an independent dynamical variable.  
As we will show in Sec.~\ref{sec:linear-mode}, the spin density shows a relaxation dynamics with its characteristic relaxation rate $\Gamma_s$.
Thus, we are now focusing on the spin hydrodynamic regime $\omega = O(\Gamma_s)$, which allows us to find balanced terms in the resulting equation of motion of the spin density.
We also note that we implicitly assume presence of a scale separation between spin and the other non-hydrodynamic modes characterized by the frequency scale $\Gamma$ as $\Gamma_s \ll \Gamma$.
This additional assumption is not what we can show in general setups, but what we expect to be true in a specific case such as the heavy-fermion mass limit.

\subsection{Derivation of constitutive relations}
To derive hydrodynamic equations with spin polarization, we need to express $\delta \Theta^\mu_{~\ha}$, $\delta q_{\ha}$, $\delta \sigma$ and $\delta J^\mu$ using our dynamical variables and background fields.
For that purpose, we require a generalized second law of local thermodynamics, which require the existence of the entropy current 
\begin{equation}
 \exists\, s^\mu (\beta,\mu, u^\mu,\mu^{\ha\hb})
  \quad \mathrm{satisfying} \quad
  (\nabla_\mu - \Gcal_\mu) s^\mu \geq 0 
  \quad \mathrm{for} \quad
  \forall\, \beta,~\mu,~u^\mu,~\mathrm{and}~ \mu^{\ha\hb}.
  \label{eq:2nd-law}
\end{equation}
Let us specify the condition following from the local second law~\eqref{eq:2nd-law}. 
By parametrizing the entropy current as $s^\mu = s u^\mu + \delta s^\mu$, 
we can express its divergence as
\begin{equation}
  (\nabla_\mu - \Gcal_\mu) s^\mu
  = u^\mu \nabla_\mu s 
  + s (\nabla_\mu u^\mu - u^\mu \Gcal_\mu)
  + (\nabla_\mu - \Gcal_\mu ) \delta s^\mu .
\end{equation}
The vital point here is that we can write the material derivative $u^\mu \nabla_\mu$ of the entropy density $s (\epsilon,n,\sigma_{\ha\hb})$ as
\begin{equation}
 u^\mu \nabla_\mu s
  = u^\mu 
  \left(
   \frac{\partial s}{\partial \epsilon} \nabla_\mu \epsilon
   + \frac{\partial s}{\partial \sigma_{\ha\hb}} 
   D_\mu \sigma_{\ha\hb}
  + \frac{\partial s}{\partial n} \nabla_\mu n
  \right)
  = \beta u^\mu 
  \left( \nabla_\mu \epsilon
  - \frac{1}{2} \mu^{\ha\hb} D_\mu \sigma_{\ha\hb}
  - \mu \nabla_\mu n
  \right),
  \label{eq:s-derivative}
\end{equation}
where we used the definition of the thermodynamic variables $\beta$, $\mu$, and $\mu^{\ha\hb}$.
Noting that the left-hand side of this equation is a scalar, we put the covariant derivative $D_\mu \sigma_{\ha\hb}$ to make the right-hand sides to be scalar too.
We can rewrite the rightmost side of the equation~\eqref{eq:s-derivative} by using the Ward-Takahashi identities with the assumed constitutive relation~\eqref{eq:general-constitutive}. 
To do so, we take contractions of Eqs.~\eqref{eq:WT} with 
$\beta^{\ha} = \beta u^{\ha} $, $\frac{1}{2} \beta \mu^{\ha\hb}$ and $\beta \mu$, which results in
\begin{align}
  \beta u^\mu \nabla_\mu \epsilon
  =& - \beta (\epsilon+p) 
  (\nabla_\mu u^\mu - u^\mu \Gcal_\mu )
  - \delta \Theta^\mu_{~\ha} 
  (D_\mu \beta^{\ha} - T^{\ha}_{~\mu\hb} \beta^{\hb} ) 
  + (\nabla_\mu - \Gcal_\mu ) (\beta \delta q^\mu)
  \nonumber
  \\
  &
  - \delta q^{\ha}
  \big[ \nabla_{\ha} \beta + u^\mu ( D_\mu \beta_{\ha}
  - \beta_{\hb} T^{\hb}_{~\mu\ha} ) 
  \big]
 \nonumber  \\
 &- \frac{1}{2} \vep^\mu_{~\ha\hb\hc} 
   (\sigma^{\hc} + \delta \sigma u^{\hc}) 
   \Rcal^{\ha\hb}_{~~\mu\hd} \beta^{\hd}
  - \beta u^\nu F_{\nu\mu} \delta J^\mu,
  \label{eq:eom-for-ds-e} 
  \\
  - \frac{1}{2}
  \beta u^\mu \mu^{\ha\hb} D_\mu \sigma_{\ha\hb} 
  =& 
  \frac{1}{2} \beta \mu^{\ha\hb} \sigma_{\ha\hb}
  ( \nabla_\mu u^\mu - u^\mu \Gcal_\mu )
  + \frac{1}{2} \beta \mu^{\ha\hb} 
  \left( 
  \delta \Theta_{\ha\hb} - \delta \Theta_{\hb\ha}
  \right) 
  \nonumber
  \\
  &+ \mu_{\ha\hb} \sigma^{\hb\mu} D_\mu \beta^{\ha}
  + ( \nabla_\mu - \Gcal_\mu) 
  (\beta \delta \sigma \mu^\mu) 
  - \frac{1}{2} \delta \sigma 
  u^{\hc} \vep^\mu_{~\ha\hb\hc} 
  D_\mu ( \beta \mu^{\ha\hb} ),
  \label{eq:eom-for-ds-sigma}
  \\
  - \beta \mu u^\mu \nabla_\mu n
  =& 
  \beta \mu n 
  (\nabla_\mu u^\mu - u^\mu \Gcal_\mu )
  - \delta J^\mu \nabla_\mu (\beta \mu)
  + (\nabla_\mu - \Gcal_\mu ) 
  (\beta \mu \delta J^\mu),
  \label{eq:eom-for-ds-n}
\end{align}
where we used 
$\delta \Theta^\mu_{~\ha} u^{\ha} = 0 = \delta q_{\ha} u^{\ha} $ to obtain the first equation.

At this point, $\delta q^{\ha}$ seems to enjoy the same status as $\delta \Theta^\mu_{~\ha}$, $\delta \sigma$ and $\delta J^\mu$, whose constitutive relation should be determined independently of other parts.
However, this is not the case: 
the constitutive relation for $\delta q_{\ha}$ is determined by that for $\Sigma^\mu_{~\ha\hb}$.
To show this, recall that the Ward-Takahashi identity for the spin current contains three constraints rather than six dynamical equations due to the totally anti-symmetric property of our spin current.
This becomes manifest by contracting Eq.~\eqref{eq:spinCurrentConservation} with $u_{\hb}$ and noticing that the resulting identity 
\begin{equation}
  \delta q^{\ha} = 
  - \frac{1}{2} \vep^{\mu\ha}_{~~\hb\hc} u^{\hb} 
  \big[
  (D_\mu- \Gcal_\mu) \sigma^{\hc} 
  + \delta \sigma D_\mu u^{\hc}
  \big] ,
  \label{eq:constraint-q}
\end{equation}
does not contain the time derivative in the right-hand side.
As a consequence, we only need to derive the constitutive relation for $\delta \Theta^\mu_{~\ha}$, $\delta \sigma$ and $\delta J^\mu$, the second of which will determine that for $\delta q_{\ha}$ in accordance with the constraint~\eqref{eq:constraint-q}.
It is thus useful to rewrite \eqref{eq:eom-for-ds-e} by substituting the constraint \eqref{eq:constraint-q} as 
\begin{equation}
 \begin{split}
  \beta u^\mu \nabla_\mu \epsilon
  =& - \beta (\epsilon +p) 
  (\nabla_\mu u^\mu - u^\mu \Gcal_\mu )
  - \delta \Theta^\mu_{~\ha} 
  (D_\mu \beta^{\ha} - T^{\ha}_{~\mu\hb} \beta^{\hb} ) 
  + (\nabla_\mu - \Gcal_\mu ) (\beta \delta q^\mu)
  \\
  &+\frac{1}{2} \vep^{\mu}_{~\ha\hb\hc} u^{\hb} 
  \big[
  (D_\mu- \Gcal_\mu) \sigma^{\hc} 
  + \delta \sigma D_\mu u^{\hc}
  \big]
  \big[ \nabla^{\ha} \beta + u^\nu ( D_\nu \beta^{\ha}
  - \beta_{\hd} T^{\hd~\ha}_{~\nu} ) 
  \big]
  \\
  &- \frac{1}{2} \vep^\mu_{~\ha\hb\hc} 
  (\sigma^{\hc} + \delta \sigma u^{\hc}) 
  \Rcal^{\ha\hb}_{~~\mu\hd} \beta^{\hd}
   - \beta u^\nu F_{\nu\mu} \delta J^\mu .
 \end{split}
\end{equation}
The result so far is exact without assuming the derivative expansion.

Let us then apply the derivative expansion to derive  relativistic spin hydrodynamic equations, which capture leading dissipative dynamics of charge densities.
For that purpose, we first need to specify our power counting scheme for the dynamical variables and background fields.
In this paper, we employ the power counting scheme defined by 
\begin{equation}
O(\partial^0) =
 \{\beta,~  u^\mu,~ \mu,~e_\mu^{~\ha}\}
 \quad \mathrm{and} \quad
 O(\partial^1) =
 \{
  \mu^{\ha\hb},~\sigma_{\ha\hb},~ 
  \omega_\mu^{~\ha\hb}\},
  \label{eq:power-counting}
\end{equation}
and derive the constitutive relations up to $O(\partial^1)$.
This allows us to keep the restricted number of terms appearing in the constitutive relations.
To demonstrate this, using Eqs.~\eqref{eq:eom-for-ds-e}-\eqref{eq:eom-for-ds-n}, we rewrite the divergence of the entropy current as
\begin{equation}
 \begin{split}
  (\nabla_\mu - \Gcal_\mu) s^\mu
  =& \big[ s - \beta (\epsilon+p-\mu n) \big] (\nabla_\mu u^\mu - u^\mu \Gcal_\mu)
  + ( \nabla_\mu - \Gcal_\mu ) 
  (\delta s^\mu + \beta \mu \delta J^\mu)
  \\
  &- \delta \Theta^{\mu}_{~\ha} \big|_{(s)}
  ( D_\mu \beta^{\ha} - T^{\ha}_{~\mu\hb} \beta^{\hb} )
  - \delta \Theta^{\mu}_{~\ha} \big|_{(a)}
  ( D_\mu \beta^{\ha} - T^{\ha}_{~\mu\hb} \beta^{\hb} 
  - \beta \mu_{\mu}^{~\ha} )
 \\
  &- \delta J^\mu 
  \big[ \nabla_\mu (\beta \mu ) - \beta^\nu F_{\mu\nu} \big]
  + \Delta S,
 \end{split}
 \label{eq:entropy-production-1}
\end{equation}
where we defined the symmetric and anti-symmetric parts of the energy-momentum tensor
\begin{equation}
 \delta \Theta^{\mu}_{~\ha} \big|_{(s)}
  \equiv \frac{1}{2} 
  ( \delta \Theta^{\mu}_{~\ha} + \delta \Theta_{\ha}^{~\mu} ) 
  \quad \mathrm{and} \quad 
  \delta \Theta^{\mu}_{~\ha} \big|_{(a)}
  \equiv \frac{1}{2} 
  ( \delta \Theta^{\mu}_{~\ha} - \delta \Theta_{\ha}^{~\mu} ).
\end{equation}
Here, we also introduced a collection of the higher-order terms $\Delta S = O(\partial^3)$, whose explicit form is given by 
\begin{align}
 \Delta S 
 \equiv &
 \frac{1}{2} \beta \mu^{\ha\hb} \sigma_{\ha\hb} 
 ( \nabla_\mu u^\mu - u^\mu \Gcal_\mu )
 + \sigma^{\mu\hb} \mu_{\hb\ha} D_\mu \beta^{\ha}
  \nonumber 
  \\
  &
  + \frac{1}{2} 
  \vep^{\mu}_{~\ha\hb\hc} 
  \left(
  u^{\hb} 
  \big[ \nabla^{\ha} \beta 
  + u^\nu ( D_\nu \beta^{\ha}
  - \beta_{\hd} T^{\hd~\ha}_{~\nu} ) 
  \big]
  (D_\mu- \Gcal_\mu) \sigma^{\hc} 
  + \sigma^{\hc} \Rcal^{\hb}_{~\ha\mu\hd} \beta^{\hd}
  \right)
  \nonumber 
  \\
  & - \frac{1}{2} \delta \sigma 
  \vep^{\mu}_{~\ha\hb\hc} u^{\hc}
  \left(
   D_\mu ( \beta \mu^{\ha\hb} )  
  + \Rcal^{\ha\hb}_{~~\mu\hd} \beta^{\hd}
  + D_\mu u^{\hb}
  \big[ \nabla^{\ha} \beta + u^\nu ( D_\nu \beta^{\ha}
  - \beta_{\hd} T^{\hd~\ha}_{~\mu} ) 
  \big]
  \right)
  \nonumber
  \\
  & + ( \nabla_\mu - \Gcal_\mu ) 
  \left( 
  \beta \delta q^\mu
  + \beta \delta \sigma \mu^\mu \right).
  \label{eq:higher-order}
\end{align}
We shall explain the reason why these terms are counted as $O(\partial^3)$. 
First of all, our power counting scheme \eqref{eq:power-counting} immediately tells us the terms in the first and second lines are the third order in derivative.
The first term in the first line, if considered, leads to a second-order correction of the Gibbs-Duhem relation as $\epsilon+p = Ts + \mu n + \frac{1}{2} \mu^{\ha\hb} \sigma_{\ha\hb}$ (see also the following discussion).
From the third line, one finds the form of $\delta \sigma$ compatible with the second law as  
\begin{equation}
 \delta \sigma
 = - T \kappa_s  \vep^{\mu}_{~\ha\hb\hc} u^{\hc}
  \left(
  D_\mu ( \beta \mu^{\ha\hb} )  
  + \Rcal^{\ha\hb}_{~~\mu\hd} \beta^{\hd}
  + D_\mu u^{\hb}
  \big[ \nabla^{\ha} \beta + u^\nu ( D_\nu \beta^{\ha}
  - \beta_{\hd} T^{\hd~\ha}_{~\mu} ) 
  \big]
  \right),
  \label{eq:delta-sigma}
\end{equation}
with a positive coefficient $\kappa_s >0$.
Equation \eqref{eq:delta-sigma} represents a diffusive transport of the spin current and its response to the curvature tensor; namely, $\kappa_s$ is identified as a spin conductivity.
Besides, this closes the constitutive relation for $\delta q_{\ha}$ thanks to the constraint~\eqref{eq:constraint-q}.
Our power counting scheme tells us $\delta \sigma = O(\partial^2)$ and $\delta q_{\ha} = O(\partial^2)$, which proves that all terms in the last two lines of Eq.~\eqref{eq:higher-order} are $O(\partial^3)$.
Note that the last term, if considered, leads to a second-order correction of the entropy current as $\delta s^\mu = - \beta \delta q^\mu - \beta \delta \sigma \mu^\mu= O(\partial^2)$. 
As a result, we find $\delta \sigma = 0$ and $\delta 
q_{\ha} = 0 $ in the first-order spin hydrodynamics.

Going back to Eq.~\eqref{eq:entropy-production-1}, one finds the entropy current $s^\mu$ and the constitutive relations for $\delta \Theta^{\mu}_{~\ha}$ and $\delta J^\mu$ compatible with the generalized second law of local thermodynamics \eqref{eq:2nd-law} in the following way.
Firstly, the first line of Eq.~\eqref{eq:entropy-production-1} vanishes by identifying
\begin{align}
 \epsilon + p = Ts +\mu n + O(\partial^2) 
 \quad \mathrm{and} \quad 
 \delta s^\mu = \beta \mu \delta J^\mu
 + O (\partial^2) ,
 \label{eq:Gibbs-Duhem}
\end{align}
where the first equation is the usual Gibbs-Duhem relation up to $O(\partial)$.
We note that both equations have second-order derivative corrections present in $\Delta S$ as we discussed above.
The second line and the first term in the third line of Eq.~\eqref{eq:entropy-production-1} will always be positive by identifying the following constitutive relations:
\begin{equation}
 \begin{split}
  \delta \Theta^{\mu}_{~\ha} \big|_{(s)}
  &= 
   - T \eta^{\mu~\nu}_{~\ha~\hb} 
  ( D_\nu \beta^{\hb} - T^{\hb}_{~\nu\hc} \beta^{\hc})
  = - \eta^{\mu~\nu}_{~\ha~\hb} 
  ( D_\nu u^{\hb} - T^{\hb}_{~\nu\hc} u^{\hc}) ,
  \\
  \delta \Theta^{\mu}_{~\ha} \big|_{(a)}
  &= - T (\etas)^{\mu~\nu}_{~\ha~\hb}
  ( D_\nu \beta^{\hb} - T^{\hb}_{~\nu\hc} \beta^{\hc} 
  - \beta \mu_\nu^{~\hb} )
  = - (\etas)^{\mu~\nu}_{~\ha~\hb}
  ( D_\nu u^{\hb} - T^{\hb}_{~\nu\hc} u^{\hc} 
  - \mu_\nu^{~\hb} ) ,
  \\
  \delta J^\mu
  &= - T \kappa_n \Delta^{\mu\nu}  
  \big[
  \nabla_\nu (\beta \mu) 
  - \beta^\rho F_{\nu\rho} 
  \big],
 \end{split}
 \label{eq:CR1}
\end{equation}
where one can decompose the viscous tensors $\eta^{\mu~\nu}_{~\ha~\hb} $ and $(\etas)^{\mu~\nu}_{~\ha~\hb}$ as
\begin{equation}
 \begin{split}
  \eta^{\mu~\nu}_{~\ha~\hb} 
  &= 2 \eta 
  \left( \frac{1}{2} 
  ( \Delta^{\mu\nu} \Delta_{\ha\hb} + \Delta^\mu_{\hb} \Delta^\nu_{\ha} )
  - \frac{1}{3} \Delta^{\mu}_{\ha} \Delta^{\nu}_{\hb}
  \right)
   + \zeta \Delta^{\mu}_{\ha} \Delta^{\nu}_{\hb}, 
  \\
  (\etas)^{\mu~\nu}_{~\ha~\hb} 
  &=  
  \frac{1}{2} 
  \etas
  (\Delta^{\mu\nu} \Delta_{\ha\hb} - \Delta^\mu_{\hb} \Delta^\nu_{\ha}).
 \end{split}
 \label{eq:viscous-tensor}
\end{equation}
Here, all four kinetic coefficients $\eta$, $\zeta$, $\etas$, and $\kappa_n$ are assumed to take positive values. 
They are shear viscosity, bulk viscosity, rotational viscosity, and flavour conductivity.
The positivity of these kinetic coefficients makes the entropy production rate a positive semidefinite quadratic form within the second-order derivative to satisfy the second law of local thermodynamics.

One can also rewrite the constitutive relation by decomposing the spin connection to the torsion-free and torsion parts as
\begin{equation}
 \begin{split}
  D_\mu u^{\ha} - T^{\ha}_{~\mu\hb} u^{\hb}
  &= \cD_\mu u^{\ha} 
  -u^{\hb} K_{\hb\mu}^{~~\ha} ,
  \\
 \end{split}
\end{equation}
where the circled derivative contains the torsion-free spin connection.
From this decomposition, one sees that the symmetric part of the energy-momentum tensor has no contribution from the torsion.
This is because the contorsion $K_{\hb\mu}^{~~\ha}$ is anti-symmetric under the 
exchange of $(\mu,\ha)$ indices so that it disappears from the constitutive relation.
Then, we obtain another expression for the constitutive relation as follows:
\begin{equation}
 \begin{split}
  \delta \Theta^{\mu}_{~\ha} \big|_{(s)}
  &= - \eta^{\mu~\nu}_{~\ha~\hb}  \cD_\nu u^{\hb} ,
  \\
 \delta \Theta^{\mu}_{~\ha} \big|_{(a)}
  &= - (\etas)^{\mu~\nu}_{~\ha~\hb}
  ( \cD_\nu u^{\hb} - u^{\hc} K_{\hc\nu}^{~~\hb} 
  - \mu_\nu^{~\hb} ).
 \end{split}
 \label{eq:CR2}
\end{equation}
Equations \eqref{eq:CR1} or \eqref{eq:CR2} give the constitutive relations for the first-order spin hydrodynamics. 
In addition to two viscosities $\eta$ and $\zeta$, we have another viscosity known as the rotational viscosity $\etas$.

In summary, we find the constitutive relations
for the first-order spin hydrodynamics 
in the torsionful curved background as follows:
\begin{equation}
 \begin{split}
  \Theta^\mu_{~\ha}
  &= \epsilon u^\mu u_{\ha} + p \Delta^{\mu}_{\ha} 
  - \eta^{\mu~\nu}_{~\ha~\hb} 
  ( D_\nu u^{\hb} - T^{\hb}_{~\nu\hc} u^{\hc})
  - (\etas)^{\mu~\nu}_{~\ha~\hb}
  ( D_\nu u^{\hb} - T^{\hb}_{~\nu\hc} u^{\hc} 
  - \mu_\nu^{~\hb} )
  \\
  &= \epsilon u^\mu u_{\ha} + p \Delta^{\mu}_{\ha} 
  - \eta^{\mu~\nu}_{~\ha~\hb}  \cD_\nu u^{\hb} 
  - (\etas)^{\mu~\nu}_{~\ha~\hb}
  ( \cD_\nu u^{\hb} - u^{\hc} K_{\hc\nu}^{~~\hb} 
  - \mu_\nu^{~\hb} ),
  \\
  \Sigma^\mu_{~\ha\hb}
  &= \vep^{\mu}_{~\ha\hb\hc} \sigma^{\hc},
  \\
  J^\mu
  &= n u^\mu -  T \kappa_n \Delta^{\mu\nu}  
  \big[
  \nabla_\nu (\beta \mu) - \beta^\rho F_{\nu\rho} 
  \big],
 \end{split}
\end{equation}
where the rank-four viscous tensors are decomposed as Eq.~\eqref{eq:viscous-tensor}.
Thus, the first-order spin hydrodynamics 
with one flavour current is equipped with three viscosities $\eta$, $\zeta$, and $\etas$
and one conductivity $\kappa_n$.
Once we know their values (and also the equation of state), the Ward-Takahashi identities form a closed set of equations for the energy-momentum,  spin, and flavor charge densities. 
This completes the phenomenological derivation of hydrodynamics with spin polarization.

A few remarks are in order here.
The present analysis uses the same strategy as Ref.~\cite{Hattori:2019lfp}, but we have some modifications that originate from the total anti-symmetry of the spin current.
This difference eliminates three dynamical spin densities attached to the boost, and, as a result, one of the transport coefficients, the boost heat conductivity, found in Ref.~\cite{Hattori:2019lfp}, disappears.
Also, the power counting scheme is a little different 
from that used in Ref.~\cite{Gallegos:2021bzp}, which makes it difficult to give a direct comparison with their constitutive relation (see the discussion at the end of Sec.~\ref{sec:linear-mode}).%
\footnote{
In principle, there can be second order derivative corrections in $\Theta_{\hat{a}\hat{b}}$, which contribute to the right-hand side of the spin-current conservation equation~\eqref{eq:spinCurrentConservation}. Those would have to be included if derivatives were our only expansion quantity, see for example~\cite{Gallegos:2021bzp}. Recall, however, that we here consider the spin relaxation rate $\Gamma_s$, which is counted on the same order as the frequency because we are considering the spin hydrodynamic regime (recall Figure~\ref{fig:energy-scale}).
} 
We here emphasize that our identification of the spin hydrodynamic regime allows a consistent truncation for the spin hydrodynamic equations.
To see this, let us consider the
Ward-Takahashi identity in the flat spacetime 
$\partial_0 \Sigma^{0}_{~\ha\hb} + \partial_i \Sigma^{0}_{~\ha\hb} = - (\Theta_{\ha\hb} - \Theta_{\hb\ha})$. 
According to our power counting scheme \eqref{eq:power-counting}, the first term on the left-hand side is counted as 
$O(\omega \partial)$ while the right-hand side will be identified as $O(\Gamma_s \partial)$ (see Sec.~\ref{sec:linear-mode}).
These are precisely balanced when we consider the spin hydrodynamic regime $\omega = O (\Gamma_s)$.
Lastly, one could try to use the novel derivative expansion scheme introduced in Ref.~\cite{Li:2020eon}, treating fluid vorticity and temperature gradients differently from other dissipative first order gradients such as shear and bulk tensors. This allows to rigorously extend the entropy-current analysis to second or higher order terms that involve vorticity and temperature gradients, producing new constraints for second or higher order transport coefficients~\cite{Li:2020eon}. The application of this scheme to the present analysis may, for example, justify (\ref{eq:delta-sigma}) more rigorously.

\section{Linear response theory for spin hydrodynamics}
\label{sec:linear}

In this section, we present the linear response theory of hydrodynamics with spin polarization.
The purpose of this section is twofold:
the first one is to specify the spectrum (or dispersion relation) of the modes described by spin hydrodynamics~\cite{Kadanoff-Martin1963}. 
We will find that the rotational viscosity controls the (gapped) relaxation rate of the spin density.
The second one is to find the field-theoretical expression for the rotational viscosity. 
Applying the linear response theory with the background (mechanical) perturbation~\cite{Green,Kubo,Nakano,Luttinger:1964zz}, we provide the Green-Kubo formula for the rotational viscosity.
Since the flavour currents (e.g., the baryon current) do not affect the dynamics of the spin density, we neglect them in the following discussion.

\subsection{Linear-mode analysis and non-hydrodynamic spin mode}
\label{sec:linear-mode}
Let us consider a spinful relativistic fluid in flat space-time, where the Ward-Takahashi identities and the constitutive relations are simplified as
\begin{equation}
 \begin{split}
   &\partial_\mu \Theta^{\mu}_{~\ha} = 0 ,
 \\
 &\partial_\mu \Sigma^{\mu}_{~\ha\hb} 
 = - ( \Theta_{\ha\hb} - \Theta_{\hb\ha} ) ,
 \\
 &\Theta^{\mu}_{~\ha} = 
 (\epsilon + p) u^\mu u_{\ha} + p \delta^\mu_{\ha}
 - \eta^{\mu~\nu}_{~\ha~\hb}
 \partial_\nu u^{\hb}
 - (\etas)^{\mu~\nu}_{~\ha~\hb} 
 ( \partial_\nu u^{\hb} - \mu_\nu^{~\hb} ),
 \\
 &\Sigma^{\mu}_{~\ha\hb} = 
 \varepsilon^{\mu}_{~\ha\hb\hc} \sigma^{\hc} .
 \label{eq:spin-hydro}
 \end{split}
\end{equation}
It is not necessary to distinguish the curved and local Lorentz indices in flat space-time, so we remove hats over the indices in the following analysis.
We perform the linear-mode analysis of the spin hydrodynamic equations \eqref{eq:spin-hydro} around the global equilibrium without background flow and spin density, that is, we consider small perturbations given by
\begin{equation}
 \epsilon (x) = \epsilon_0 + \delta \epsilon (x), \quad 
  v^i (x) =  0 + \delta v^i (x), \quad 
  \sigma^{\ha} (x) = 0 + \delta \sigma^{\ha} (x),
  \label{eq:liner-fluctuation}
\end{equation}
and keep only $O(\delta)$-terms in the following analysis.
In this expansion, the four-velocity and spin densities are given by 
$u^\mu = (1,\delta v^i)^t + O(\delta^2)$ and 
$\delta \sigma^{a} = (0,\delta \sigma^{i})^t + O (\delta^2)$, and the fluctuations of all other quantities (such as the pressure) are expressed by those in Eq.~\eqref{eq:liner-fluctuation}.
Introducing $\delta \pi_{i} \equiv \Theta^0_{~i} = (\epsilon_0 + p_0) \delta v_{i}$, one finds that the linearized equations of motion become
\begin{equation}
 \begin{split}
  0 &= \partial_0 \delta \epsilon + \partial_i \delta \pi^i,
  \\
  0 &= \partial_0 \delta \pi_i + c_s^2 \partial_i \delta \epsilon 
  - \gamma_\para \partial_i \partial^j \delta \pi_j
  - ( \gamma_\perp + \gamma_{s} ) (\delta_i^j \bnab^2 - \partial_i \partial^j)
  \delta \pi_j
  + \frac{1}{2} \Gamma_s \vep_{0ijk} \partial^j \delta \sigma^k,
  \\
 0 &=
  \partial_0 \delta \sigma_i
  + \Gamma_s \delta \sigma_i 
  + 2 \gamma_s \vep_{0ijk} \partial^j \delta \pi^k,
  \\
 \end{split}
\end{equation}
where we introduced a set of static/kinetic coefficients as
\begin{equation}\label{spinsus}
 \begin{split}
  &c_s^2 \equiv \frac{\partial p}{\partial \epsilon}, \quad 
  \gamma_{\para} \equiv \frac{1}{\epsilon_0 + p_0} 
  \left( \zeta + \frac{4}{3} \eta \right)
  , \quad 
  \gamma_{\perp} \equiv \frac{\eta}{\epsilon_0 + p_0}
  , 
  \\
  &\chi_s \delta_{ij}
  \equiv \frac{\partial \sigma_i}{\partial \mu^j}
  , \quad 
  \gamma_s \equiv 
  \frac{\etas}{2(\epsilon_0+p_0)}, \quad 
  \Gamma_s \equiv \frac{2\etas}{\chi_s} .
 \end{split}
\end{equation}
The $\chi_s$ is the spin susceptibility, and $c_s$ is the speed of sound.
We work in the Fourier space with
$\delta \tilde{\Ocal} (k) 
\equiv \ds{\int} \diff^4 x\,\, \rme^{\rmi \omega t - \rmi \bk \cdot \bx} \delta \Ocal (x)$, and we choose the momentum direction to be along the $z$-direction,
$\bk = (0,0,k)$.
Then, the linearized equations of motion become the block diagonal form as 
\begin{equation}
 \left(
 \begin{array}{c|c}
  A_{\para}^{3 \times 3} & O \\ \hline 
   O & A_{\perp}^{4 \times 4} 
 \end{array}
 \right)
 \delta \vec{c}  
 = 0
 \with
 \delta \vec{c} \equiv 
  \begin{pmatrix}
   \delta \tilep &
   \delta \tilpi_z &
   \delta \tilsigma_z &
   \delta \tilpi_x &
   \delta \tilsigma_y &
   \delta \tilpi_y &
   \delta \tilsigma_x 
  \end{pmatrix}^t,
\end{equation}
where we introduce the parallel and perpendicular matrices as
\begin{equation}
 \begin{split}
  A_{\para}^{3 \times 3}
  &\equiv 
  \begin{pmatrix}
   - \rmi \omega & \rmi |\bk| & 0  \\
   \rmi c_s^2 |\bk| & - \rmi \omega + \gamma_{\para} \bk^2 & 0  \\
   0 & 0 & - \rmi \omega + \Gamma_s 
  \end{pmatrix}, 
  \\
  A_{\perp}^{4 \times 4}
 &\equiv 
  \begin{pmatrix}
   -\rmi \omega + (\gamma_\perp + \gamma_s) \bk^2 &  \frac{\rmi}{2} \Gamma_s |\bk| & 0 & 0 \\
   - 2 \rmi \gamma_s |\bk|  & - \rmi \omega + \Gamma_s & 0 & 0 
   \\
   0 & 0 & -\rmi \omega + (\gamma_\perp + \gamma_s) \bk^2 & \frac{\rmi}{2} \Gamma_s |\bk| \\
   0 & 0 & - 2 \rmi \gamma_s |\bk| & - \rmi \omega + \Gamma_s
  \end{pmatrix}. 
 \end{split}
 \label{eq:block-matrix}
\end{equation}
Solving the characteristic equations, 
$\det A_{\para}^{3 \times 3} = 0$ and 
$\det A_{\perp}^{4 \times 4} =0$, we find the dispersion relations for the modes in our spin hydrodynamics.

\begin{figure}[t]
 \centering
 \includegraphics[width=1.0\linewidth]{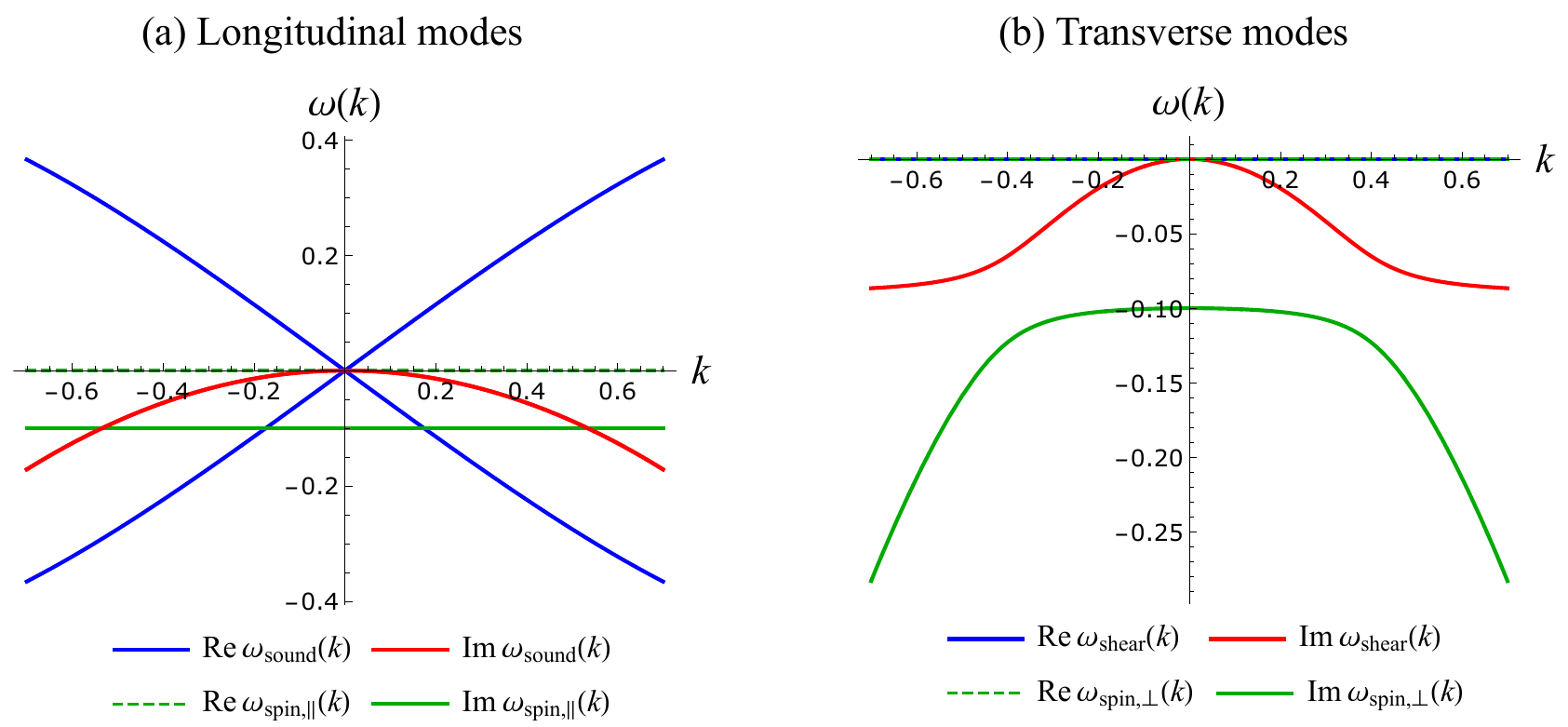}
 \caption{Dispersion relations for (a) longitudinal modes and (b) transverse modes, with the parameters
 $(c_s,\gamma_{\para},\gamma_{\perp},\Gamma_s,\gamma_s) = (1/\sqrt{3},0.7,0.5,0.1,0.05) $.
}
 \label{fig:DR-spin-hydro}
\end{figure}

The longitudinal modes, described by $A^{3\times 3}_{\para}$, decompose into the decoupled sound and spin modes, whose dispersion relations read
\begin{equation}
 \begin{cases}
  \bullet~
  \mathrm{One~pair~of~sound~modes}:~
  \omega_{\mathrm{sound}} (\bk)
  = \pm c_s |\bk| - \dfrac{\rmi}{2} \gamma_{\para} \bk^2 + O(\bk^3),
  \\
  \bullet~
  \mathrm{One~longitudinal~spin~mode}:~
  \omega_{\mathrm{spin,\para}} (\bk)
  = - \rmi \Gamma_s .
 \end{cases}
 \label{eq:Dispersion1}
\end{equation}
It is clear that we have the relaxational longitudinal spin mode in addition to the gapless sound mode (see Figure~\ref{fig:DR-spin-hydro}(a)).
In contrast, the transverse momentum (shear) and spin modes are coupled together, which gives rise to two pairs of the following dispersion relations (see Figure~\ref{fig:DR-spin-hydro}(b)):
\begin{equation}
 \begin{split}
  \omega_{\mathrm{shear}} (\bk)
  &= - \dfrac{\rmi \Gamma_s + \rmi (\gamma_{\perp} + \gamma_s) \bk^2 - \rmi \sqrt{\Gamma_s^2 
  - 2 \Gamma_s(\gamma_\perp - \gamma_s ) \bk^2
  + (\gamma_\perp + \gamma_s)^2 \bk^4 }}{2},
  \\
  \omega_{\mathrm{spin,\perp}} (\bk)
  &= - \dfrac{\rmi \Gamma_s + \rmi (\gamma_{\perp} + \gamma_s) \bk^2 + \rmi \sqrt{\Gamma_s^2 
  - 2 \Gamma_s (\gamma_\perp - \gamma_s ) \bk^2
  + (\gamma_\perp + \gamma_s)^2 \bk^4 }}{2}.
 \end{split}
 \label{eq:Dispersion-exact}
\end{equation}
Here, we call each branch as the shear and the transverse spin mode, because of their low-momentum behaviors 
\begin{equation}
 \begin{cases}
 \bullet~
 \mathrm{Two~shear~modes}:~
  \omega_{\mathrm{shear}} (\bk)
  = - \rmi \gamma_{\perp} \bk^2 + O (\bk^4),
  \\
  \bullet~
  \mathrm{Two~transverse~spin~modes}:~
  \omega_{\mathrm{spin,\perp}} (\bk)
  = - \rmi \Gamma_s - \rmi \gamma_s \bk^2 
  + O (\bk^4).
 \end{cases}
 \label{eq:Dispersion2}
\end{equation}
From the low-momentum behaviors in \eqref{eq:Dispersion1} and \eqref{eq:Dispersion2}, we identify three non-hydrodynamic (gapped) spin modes, in addition to the hydrodynamic (gapless) sound and shear modes.
All spin modes show the relaxation behavior with the characteristic time scale of $\Gamma_s^{-1}$.
This non-hydrodynamic behavior is originated from the fact that the spin angular momentum is not a conserved quantity. 
The relaxation of spin modes persists until the source term in the spin equation of motion vanishes, namely when the spin potential $\mu^{ab}$ coincides with the fluid vorticity when it reaches the local equilibrium.
In this strict hydrodynamic regime of $\omega \ll \Gamma_s$,
we thus find only the gapless sound and shear modes, as expected in the conventional relativistic hydrodynamics.

On the other hand, in the regime of our spin hydrodynamics ($\omega \sim \Gamma_s$), three components of spin density start to behave as independent dynamical modes.
In contrast to the decoupled longitudinal spin mode, the transverse spin modes feature an interesting coupled dynamics with the transverse shear mode.
In fact, in high momentum region,
the shear and transverse spin branches in the dispersion relations behave as
\begin{equation}
 \omega_{\mathrm{shear}} (\bk) 
 \sim 
 - \frac{\rmi \gamma_\perp \Gamma_s}{\gamma_\perp + \gamma_s}
 \quad \mathrm{and} \quad 
 \omega_{\mathrm{spin,\perp}} (\bk) 
 \sim 
 - \rmi ( \gamma_\perp + \gamma_s ) \bk^2.
\end{equation}
It is notable that they interchange their momentum dependence between the low and the high momentum regimes: the shear branch approaches to a constant relaxation rate independent of momentum, while the spin branch shows a quadratic dependence on momentum (see Figure~\ref{fig:DR-spin-hydro}(b)). 
At the cross-over momentum scale, we see the level repulsion phenomenon between the two branches.
We should also point out that the eigenvector of each branch is a linear combination of $\delta \pi_{x}$ and $\delta \sigma_y$ (or $\delta \pi_{y}$ and $\delta \sigma_x$), due to the off-diagonal mixing terms in the matrix $A_{\perp}^{4 \times 4}$ in Eq.~\eqref{eq:block-matrix}.
This mixing is a result of the fact that the transverse velocity gradient $\partial_z \delta v_x$ is a linear combination of shear tensor and vorticity, and the former couples to momentum fluctuation while the latter couples to the spin density.
In particular, the strict hydrodynamic shear mode in low momentum is described by the energy-momentum tensor $T^\mu_{~\ha}$ defined in Eq.~\eqref{eq:EM-T}, which is a linear combination of our energy-momentum tensor $\Theta^{\mu}_{~\ha}$ and the spin density.
In fact, the flat space-time limit of Eq.~\eqref{eq:EM-T} gives us
\begin{equation}
 \delta \tilT^0_{~i}
 = \delta \tilpi_i 
 + \frac{\rmi}{2} \vep_{0ijk} k^k \delta \sigma^j,
 \label{eq:diagonalized}
\end{equation}
and we find that the transverse fluctuation, e.g., $\delta \tilT^0_{~x}$, is one of the eigenvectors in low momentum regime satisfying 
\begin{equation}
 0 = \big[ - \rmi \omega + \gamma_\perp k^2 \big]
  \delta \tilT^0_{~x} + O (\bk^3).
\end{equation}
This shows that $T^0_{~\ha}$ describes the gapless hydrodynamic modes in the strict hydrodynamic regime. 
However, we emphasize that we need $\Theta^\mu_{~\ha}$ to describe the dynamics of spin polarization, because its anti-symmetric part appears in the equation of motion for spin polarization.

The zero-momentum relaxation rate of the spin modes is controlled by $\Gamma_s = 2 \etas/\chi_s$, and thus, the rotational viscosity $\etas$ is essentially the spin relaxation rate.
Note that the finite-momentum part of the spin relaxation rate is different between the longitudinal and transverse components:
the former does not depend on momentum, while the latter is given also by the rotational viscosity $\etas$ as 
$\gamma_s \bk^2 = \frac{\etas}{2(\epsilon_0+p_0)} \bk^2$.
If we consider the next-leading-order correction of the spin current, the relaxation rate for the longitudinal spin density also shows a momentum dependence.
While the finite-momentum dependence of the spin relaxation is an interesting subject, we focus on the zero-momentum behavior in the remainder of this paper.
We also note that the parameter set used in Figure~\ref{fig:DR-spin-hydro} (and also in Figure~\ref{fig:energy-scale}) is chosen to have a well-defined spin hydrodynamic regime.  This is manifest in our choice of small $\Gamma_s$, compared to the scale where hydrodynamic gradient expansion breaks down, i.e., $\omega_c\sim c_s^2/\gamma_\parallel$. 
This is a required assumption that ensures
the existence of the spin hydrodynamic regime (or the spin hydro$+$ regime in Figure \ref{fig:energy-scale}), in which only spin modes appear as an additional non-hydrodynamic mode in the time scale much larger than that of other non-hydrodynamic modes. 
This may be justified when, e.g., there exist sufficiently heavy fermions in the system.

Let us translate the above results into field-theoretic language. 
For that purpose, we introduce the retarded Green's function for arbitrary operators $\hA$ and $\hB$ as
\begin{equation}
 \tilG_{\ret}^{A,B} (\omega,\bk) 
  \equiv \int \diff^4 x\,\, \rme^{\rmi \omega t - \rmi \bk \cdot \bx}
  \,\rmi \theta (t)
  \baverage{[\hA(t,\bx), \hB(0,\bzero)]}_{\mathrm{eq}}, 
  \label{eq:retarded-Green}
\end{equation}
where $\average{\hOcal}_{\mathrm{eq}}$ denotes the expectation value in homogeneous thermal equilibrium.
Then, the dispersion relation~\eqref{eq:Dispersion1} and ~\eqref{eq:Dispersion2} implies that the retarded Green's function for the spin density $G_{\ret}^{\sigma^i \sigma^j} (\omega, \bk = 0)$ has a pole at the location $\omega_{\mathrm{spin}} (\bk=0) = - \rmi \Gamma_s$.
Besides, taking into account that the zero-frequency limit of the retarded Green's function gives the spin susceptibility 
\begin{equation}
 \lim_{\bk \to 0} \tilG_{\ret}^{\sigma_i \sigma_j}
  (\omega = 0 ,\bk)
  = \chi_s \delta^{ij},
\end{equation}
we identify the low frequency-momentum behavior of the retarded Green's function for spin densities as 
\begin{equation}
  \tilG_{\ret}^{\sigma^i \sigma^j} (\omega, \bk )
   = \frac{\rmi \chi_s \Gamma_s + \cdots}   {\omega + \rmi \Gamma_s + O (\bk^2)} \delta^{ij} ,
   \label{eq:Retarded-spin-spin}
 \end{equation}
where the ellipsis stands for higher-order corrections with respect to frequency and momentum, which is beyond our first-order spin hydrodynamics.
As a result, we can directly extract the value of the spin relaxation rate $\Gamma_s$, if we identify the location of the pole in the retarded spin-spin correlator.
This give one direct way to evaluate the spin relaxation rate from the correlation function in quantum field theory.

Before closing this section, we 
comment on our power counting scheme.
Noting that the equation of motion for the spin density is given by Eq.~\eqref{eq:spinCurrentConservation}, one might worry whether our constitutive relation is consistent with (naive) derivative expansion scheme.
This is a natural concern, because the left-hand side of Eq.~\eqref{eq:spinCurrentConservation} is of second order in derivatives, while the right-hand side is of first order.
However, the crucial point is that we need to distinguish spatial derivatives from time derivatives in the spin hydrodynamic regime we are working on, since
our frequency scale is implicitly assumed to be
$\omega \gtrsim \Gamma_s$, and therefore, time derivative should not be counted equally to spatial derivative.
Then, there is no discrepancy in the number of derivatives between both sides of 
Eq.~\eqref{eq:spinCurrentConservation}: 
they are both $O(\Gamma_s \partial)$, where $\partial$ denotes a spatial derivative.
This point will also be important when we check the consistency between the linear-mode analysis and the Green-Kubo formula in the next section.

\subsection{Green-Kubo formula in the first-order formalism}
As shown in the previous section, the rotational viscosity $\etas$ gives the relaxation rate of spin polarization.
Our primary question is how to evaluate it from an underlying QFT such as QCD.
We formulate in this section the linear-response theory based on the first-order formalism, which gives the Green-Kubo formula to evaluate the spin relaxation rate from the field-theory Green's functions.

First of all, we note that the spin density is a non-hydrodynamic mode with a characteristic relaxation rate 
$\Gamma_s$, and therefore, it disappears in the strict hydrodynamic regime $\omega \ll \Gamma_s$ in our setup, where the local equilibrium state does not support vorticity.
Thus, when writing the equations of spin hydrodynamics, we implicitly assume that we are working on the frequency scale given by $\omega \gtrsim \Gamma_s$ (recall Figure~\ref{fig:energy-scale}).
In other words, depending on the frequency scale of interest to us, the constitutive relation of the spin hydrodynamics is further reduced to be the usual hydrodynamics as
\begin{equation}
\delta \Theta^{\mu}_{~\ha} \big|_{(a)}
  = 
  \begin{cases}
  - (\etas)^{\mu~\nu}_{~\ha~\hb}
  ( \cD_\nu u^{\hb} - u^{\hc} K_{\hc\nu}^{~~\hb} 
  - \mu_\nu^{~\hb} )
  \quad \mathrm{when} \quad
  \Gamma_s \ll \omega \ll \Gamma,
  \\ 
  0 
  \hspace{154pt} \mathrm{when} \quad
  \omega \ll \Gamma_s,
  \end{cases}
\end{equation}
where $\Gamma$ represents the relaxation scale of other non-hydrodynamic modes.
Here, we specified the constitutive relation in the true hydrodynamic regime $\omega \ll \Gamma_s$ by 
equating the source term of the spin equation of motion with zero.
As a result, the small frequency limit which we will use below is not the strict $\omega \to 0$ limit, but should be regarded as a constrained limit
$\Gamma_s \ll \omega \ll \Gamma$.
This is in sharp contrast to the usual $\omega \to 0$ limit used in the Green-Kubo formulae for other transport coefficients for conserved charges.

We then introduce a mechanical perturbation~\cite{Luttinger:1964zz}, and identify the rotational viscosity from the linear response of the system to the background perturbation.
The basic strategy is to compute the perturbation of currents in first order of the background torsion, and to match the result with the constitutive relation in our spin hydrodynamics, \eqref{eq:CR1} or \eqref{eq:CR2}.
From the first equation of \eqref{eq:CR2}, one sees that the bulk and shear viscosities are captured by the torsion-free spin connection, which results in the usual Green-Kubo formula for them.
Thus, we focus on the rotational viscosity in the following discussion.

The vital point for the subsequent discussion is 
to identify Eqs.~\eqref{eq:CR1} or \eqref{eq:CR2} as 
a non-equilibrium expectation value of the quantum operator 
$\average{ \delta \hTheta^\mu_{~\ha}}$ in a torsionful curved background.
We here use the vierbein and contorsion as independent sources, 
while it is also possible to use the vierbein and spin connection, instead.
Similarly to the previous section, we assume that the system is initially in homogeneous thermal equilibrium with $\beta = \mathrm{const.}$, $u^{\ha} = (1,\bm{0})^t$, and $\mu_\mu^{~\ha} = 0 $.
On the time scale of $\omega\gg \Gamma_s$, the spin chemical potential $\mu_\mu^{~\ha}$ remains zero even after introducing background perturbations, as we explained in the previous paragraph.
This simplifies the constitutive relation for the anti-symmetric part of the energy-momentum tensor as
\begin{equation}
 \begin{split}
  \average{ \delta \hTheta^\mu_{~\ha} \big|_{(a)} }_{e,K}
  &= - \frac{1}{2} \etas 
  ( \Delta^{\mu\nu} \Delta_{\ha\hb} 
  - \Delta^\mu_{\hb} \Delta_{\ha}^\nu ) 
  \big[ \comega_{\nu~\hzero}^{~\hb} (e) - K_{\hzero\nu}^{~~\hb} \big].
 \end{split}
 \label{eq:Theta-curved1}
\end{equation}
This equation serves as a basis to obtain the Green-Kubo formula for the rotational viscosity $\etas$.
Below, we present two ways to compute the rotational viscosity, whose equivalence will be shown by the Ward-Takahashi identity.

Firstly, we turn off the background vierbein by putting $e_\mu^{~\ha} = \delta_\mu^{\ha}$, while keeping the non-vanishing contorsion, so that one has only the second term in Eq.~\eqref{eq:Theta-curved1}.
On the other hand, we can also compute the same $\average{\delta \hTheta^\mu_{~\ha} }_{e.K}$ in a different manner based on QFT.
To describe this in the underlying quantum theory, 
we start from the initial density operator at past infinity $t =- \infty$ given by the equilibrium Gibbs distribution
\begin{equation}
 \hrhoeq = \frac{1}{Z (\beta)} \rme^{-\beta \hH}
  \with
  Z (\beta) \equiv \Tr \rme^{-\beta \hH},
\end{equation}
where $\hH$ denotes the Hamiltonian of, e.g., QCD in the absence of background fields.
Then, noting that Eq.~\eqref{eq:Theta-curved1} indicates that the contorsion 
$K_{\hzero\nu}^{~~\hb}$ plays a role of a relevant background field, we perturb the system by adiabatically turning on the small contorsion and compute the induced response of the anti-symmetric part of $\average{\hTheta^{\mu}_{~\ha}}$.
Since the contorsion couples to the spin current in the microscopic field theory, one can describe the situation by considering the time-evolution generated by a time-dependent Hamiltonian
\begin{equation}
 \hH_{\mathrm{tot}} (t) = \hH + \hH_{\ext} (t)
  \with
  \hH_{\ext} (t)
  \equiv 
  \frac{1}{2}
  \int \diff^3 x K_{\hzero\nu}^{~~\hb} (t,\bx) 
  \hSigma^{\hzero\nu}_{~~\hb} (t,\bx).
  \label{eq:time-dep-H1}
\end{equation}
We shall then evaluate the first-order perturbation induced by
$\hH_{\mathrm{ext}} (t)$.
A usual computation easily gives the energy-momentum tensor $\average{\hTheta^\mu_{~\ha} (x)}_{K}$ as
\begin{align}
  \average{\hTheta^{\mu}_{~\ha} (x)|_{(a)}}_K
  &= \average{\hTheta^{\mu}_{~\ha} (x)|_{(a)}}_{\eq}
  - \frac{\rmi}{2} \int_{-\infty}^{\infty} \diff t' \diff^3 x'
  \theta (t-t')
  \baverage{[\hTheta^{\mu}_{~\ha} (x)|_{(a)}, \hSigma^{\hzero\nu}_{~~\hb} (x') ]}_{\eq} 
  K_{\hzero\nu}^{~~\hb} (x')
 \nonumber \\
  &= \average{\hTheta^{\mu}_{~\ha} (x)|_{(a)}}_{\eq}
  - \frac{1}{2} \int_{-\infty}^{\infty} \diff t' \diff^3 x'
  G_{\ret}^{\hTheta^{\mu}_{~\ha}|_{(a)},\Sigma^{\hzero\nu}_{~~\hb}} (x-x')
  K_{\hzero\nu}^{~~\hb} (x'),
\end{align}
where we use the definition of the retarded Green's function~\eqref{eq:retarded-Green} in the second line.
The first term with the subscript ``eq'' on the right-hand side represents the equilibrium expectation value in the absence of the background contorsion.
Thus, after applying the Markov approximation, which essentially means a small frequency limit, we find the linear response result for 
$\average{\delta \hTheta^{\mu}_{~\ha} (x)|_{(a)}}
= \average{\hTheta^{\mu}_{~\ha} (x)|_{(a)}}_K
- \average{\hTheta^{\mu}_{~\ha} (x)|_{(a)}}_{\eq}$ as
\begin{equation}
 \begin{split}
  \average{\delta \hTheta^{\mu}_{~\ha} (x)|_{(a)}}
  \simeq 
  - \frac{1}{2} \lim_{\Gamma_s \ll \omega \ll \Gamma} 
  \lim_{\bk \to 0} 
  \tilG_{\ret}^{\Theta^{\mu}_{~\ha}|_{(a)},\Sigma^{\hzero\nu}_{~~\hb}} 
  (\omega,\bk) 
  K_{\hzero\nu}^{~~\hb} (x)\, .
 \end{split}
 \label{eq:Linear-response}
\end{equation}
Note that we use the constrained limit,
$\ds{\lim_{\Gamma_s \ll \omega \ll \Gamma}}$, that was discussed previously, since in the strict $\omega\to 0$ limit, both sides are expected to vanish due to relaxation dynamics of spin density.
By comparing this result with Eq.~\eqref{eq:Theta-curved1}, 
we find the rotational viscosity given by 
\begin{equation}
  \etas = 
  - \lim_{\Gamma_s \ll \omega \ll \Gamma}
  \lim_{\bk \to 0}
  \tilG_{\ret}^{\Theta^{x}_{~\hy}|_{(a)},\Sigma^{\hzero x}_{~~\hy}} 
  (\omega,\bk) .
  \label{eq:Green-Kubo-1}
\end{equation}
This gives one Green-Kubo formula for the rotational viscosity.

As we mentioned previously, there is another way to compute the rotational viscosity.
This is because Eq.~\eqref{eq:Theta-curved1} also involves the vierbein through the torsion-free spin connection $\comega_{\nu~\hzero}^{~\hb} (e)$.
Thus, considering the linear response induced by the anti-symmetric part of the vierbein instead of the contorsion, we can also compute the rotational viscosity. 
However, we need to be careful at this point because the operator coupled to the vierbein is not $\hTheta^{\mu}_{~\ha}$ but $\hT^{\mu}_{~\ha}$ defined in Eq.~\eqref{eq:EM-T} when we treat the vierbein and contorsion as independent sources.
This second way to compute the rotational viscosity suggests that we can use the torsion-free background to compute the rotational viscosity, but the identification of $\Theta^\mu_{~\ha}$ will be a little complicated (see also Appendix~\ref{sec:metric}).

We can, however, derive the same result while avoiding such a direct computation with the vierbein perturbation.
To do so, we simply need to use the Ward-Takahashi identity for the spin current.
In flat space-time, we find the following identity in the Fourier space:
\begin{equation}
  - \rmi \omega 
  \tilSigma^{\hzero x}_{~~\hy} 
  (\omega,\bk)
  + \rmi k 
  \tilSigma^{zx}_{~~\hy} (\omega,\bk)
  = - 2 \tilTheta^{x}_{~\hy} \big|_{(a)} (\omega,\bk).
  \label{eq:WT-Fourier}
\end{equation}
Substituting this relation into Eq.~\eqref{eq:Green-Kubo-1}, we obtain another Green-Kubo formula for the rotational viscosity: 
\begin{equation}
  \etas = 
   2 \lim_{\Gamma_s \ll \omega \ll \Gamma} 
  \lim_{\bk \to 0}
  \frac{1}{\omega}
  \im \tilG_{\ret}^{\Theta^{x}_{~\hy}|_{(a)},\Theta^{x}_{~\hy}|_{(a)}} 
  (\omega,\bk) .
  \label{eq:Green-Kubo-2}
\end{equation}
This looks more familiar because it takes a similar form to the Green-Kubo formula for the shear and bulk viscosities.

We can show that the constrained limit $\ds{\lim_{\Gamma_s \ll \omega \ll \Gamma}}$ is indeed necessary in these Green-Kubo formulas.
To see this, using the Ward-Takahashi identity \eqref{eq:WT-Fourier},
we can replace $\Theta^{x}_{~\hy}$ in the first Green-Kubo formula  \eqref{eq:Green-Kubo-1} with the spin density, which leads to
another expression for the rotational viscosity:
\begin{equation}
  \etas = 
  \frac{1}{2}
  \lim_{\Gamma_s \ll \omega \ll \Gamma}  \lim_{\bk \to 0}
  \omega
    \im \tilG_{\ret}^{\Sigma^{\hzero x}_{~~\hy},\Sigma^{\hzero x}_{~~\hy} }
  (\omega,\bk) .
  \label{eq:Green-Kubo-3}
\end{equation}
This equation is consistent with
the result of the linear-mode analysis given in Eq.~\eqref{eq:Retarded-spin-spin} 
(recall $\sigma^z = \Sigma^{\hzero x}_{~~\hy}$ so that $\tilG_{\ret}^{\sigma^z \sigma^z} (\omega,\bk) = \tilG_{\ret}^{\Sigma^{\hzero x}_{~~\hy},\Sigma^{\hzero x}_{~~\hy} } (\omega,\bk)$).
Indeed, the constrained limit of Eq.~\eqref{eq:Retarded-spin-spin} leads to 
\begin{equation}
\omega \tilG_{\ret}^{\sigma^i \sigma^j} (\omega, \bk =0 )
 = \frac{\rmi \chi_s \omega \Gamma_s + O(\omega^2)} {\omega + \rmi \Gamma_s } \delta^{ij} 
 \xrightarrow{\Gamma_s \ll \omega \ll \Gamma}
 \frac{\rmi \chi_s \omega \Gamma_s}{\omega}
 = 2 \rmi \eta_s,
\end{equation}
which agrees with the linear response result~\eqref{eq:Green-Kubo-3}.
On the other hand, Eq.~\eqref{eq:Retarded-spin-spin} tells us that 
the strict $\omega \to 0$ limit of $\omega \im \tilG_{\ret}^{\Sigma^{\hzero x}_{~~\hy},\Sigma^{\hzero x}_{~~\hy} }  (\omega,\bk)$ does not give the rotational viscosity, because it simply vanishes.

While the above Green-Kubo formula utilizes the constrained limit, the strict $\omega \to 0$ limit can also provide yet another formula for the rotational viscosity.
Taking the imaginary part of Eq.~\eqref{eq:Retarded-spin-spin}, it is straightforward to obtain
\begin{equation}
  \frac{\chi_s^2}{2\etas}
 = \lim_{\omega \to 0} \frac{1}{\omega}
 \im 
 \tilG_{\ret}^{\Sigma^{\hzero x}_{~~\hy},\Sigma^{\hzero x}_{~~\hy} } (\omega,\bk=0).
  \label{eq:Retarded-spin-spin2}
\end{equation}
We emphasize that $\ds{\lim_{\omega \to 0}}$ in this equation is the usual zero-frequency limit.
Interestingly, the rotational viscosity $\etas$ appears in the denominator of the equation, in contrast to the other Green-Kubo formula that we derive in this section. This can be understood from the requirement $\omega\ll \Gamma_s$ in this limit, which probes infrared singularities that should be regularized by a finite $\Gamma_s$.
The Eq.~\eqref{eq:Retarded-spin-spin2} may provide another practical way to evaluate the rotational viscosity in quantum field theory.

\section{Summary and Outlook}
\label{sec:Summary}
In this paper, we addressed several theoretical issues in relativistic hydrodynamics with spin polarization.
First, we clarified the definition of the energy-momentum tensor and spin current for quantum field theories with Dirac fermions (such as QED and QCD) based on the first-order formalism in a background geometry with torsion.
The resulting spin current is totally anti-symmetric with respect to its three indices, and we performed the consistent entropy current analysis to derive the first-order constitutive relations in the torsionful background. 
The total anti-symmetry of the spin current leads to three dynamical degrees of freedom arising from spin polarization, and we find one kinetic coefficient known as the rotational viscosity, which controls the relaxation rate for the spin density.
Second, we presented a number of ways to evaluate the spin relaxation rate, or equivalently the rotational viscosity, from the retarded correlation functions of spin-related operators; the several Green-Kubo formulae for the rotational viscosity are derived with the help of mechanical perturbations described by the time-dependent Hamiltonian.
We also point out a subtlety of the low-frequency limit in the obtained Green-Kubo formula, which originates from the non-conservation of spin angular momentum that is inherent in relativistic spin hydrodynamics.

Although we presented a rigorous framework for the first-order relativistic spin hydrodynamics, we must also point out the underlying assumptions for its validity, which have not been clearly explained in the previous literature. 
As was already pointed out in the old paper by Martin, Parodi, and Pershan~\cite{Martin1972}, we cannot model-independently assume that the relaxational spin mode is in the near hydrodynamic regime.
In other words, we cannot always find the regime for the spin hydrodynamics we discuss, which is defined by $\Gamma_s \ll \omega \ll \Gamma$, while we use this assumption to derive the Green-Kubo formula for the rotational viscosity.
This is in sharp contrast to the strict hydrodynamic modes whose gapless property is protected by conservation laws, so that the presence of the hydrodynamic regime is always guaranteed in the $\bk \to 0$ limit.  
There is a certain similarity between spin hydrodynamics and M{\"u}ller-Israel-Stewart theory of second-order relativistic hydrodynamics~\cite{Muller:1967zza,Israel:1976tn,Israel:1976-2,Israel:1979wp} in that both introduce additional degrees of freedom. 
Neither of these two theories could be a 
universal effective description unless the additional degrees of freedom are parametrically slow. 
However, when an approximate spin symmetry emerges by considering, e.g., a large mass limit for fermions (e.g., charm quarks in QCD, and protons and/or electrons in QED), 
$\Gamma_s^{\mathrm{heavy}}$ for heavy-fermion spin is expected to be parametrically smaller than $\Gamma$ for other non-hydrodynamic modes, which guarantees the existence of the spin hydrodynamic regime.

As an outlook let us point out a few research directions along the lines of the present paper.
The most important one is to evaluate the rotational viscosity (or the spin relaxation rate) for QCD or gauge theories. 
The Green-Kubo formula for the rotational viscosity derived in this paper enables us to compute the spin relaxation rate in the underlying quantum field theory.
For that purpose, similarly to the computation of the shear viscosity,
we may rely on thermal perturbation theory for the weakly-coupled regime of QCD~\cite{Arnold:2000dr,Arnold:2003zc}, or the AdS/CFT correspondence for the strongly-coupled limit of QCD-like theories~\cite{Policastro:2001yc,Policastro:2002se}. 
The results would have direct implications for the theoretical study of spin polarization in the QGP created in relativistic heavy-ion collision experiments.

Our analysis in this paper employs the entropy-current analysis to derive the constitutive relations. 
Another interesting direction is to explore the non-equilibrium statistical operator method~\cite{Nakajima,Mori1,McLennan,McLennan1,Zubarev1979,Zubarev1,Zubarev2,Kawasaki-Gunton,Huang:2011dc,Sasa2014,Becattini:2014yxa,Hayata:2015lga,Becattini:2019dxo,Hongo:2020qpv}. 
This will provide a complementary starting point to discuss spin transport with the help of thermal perturbation theory, which describes the relaxation process based on density operators~(see, e.g., Ref.~\cite{Hu:2021lnx} for a recent attempt to derive the Green-Kubo formula with the statistical operator method). 
Besides, it will also be interesting to investigate possible anomalous transport phenomena in the spin transport, with the help of the path-integral formulation of the local thermodynamic functional~\cite{Hongo:2016mqm}. 
In fact, non-dissipative anomalous currents such as the chiral magnetic and vortical effects~\cite{Fukushima:2008xe,Erdmenger:2008rm,Banerjee:2008th,Son:2009tf,Landsteiner:2011cp,Landsteiner:2016led,Kharzeev:2020jxw}
are captured by the local thermodynamic functional, or equivalently called the hydrostatic partition function~\cite{Banerjee:2012iz,Jensen:2012jh,Jensen:2012jy,Jensen:2013kka,Golkar:2015oxw,Chowdhury:2016cmh,Glorioso:2017lcn,Manes:2018llx,Hongo:2019rbd,Manes:2019fyw}. 
Guided by the result obtained in Ref.~\cite{Hongo:2016mqm}, an extra amount of spin density, which corresponds to the deviation of the spin potential from its equilibrium value of thermal vorticity, would lead to the presence of \textit{thermal torsion} emerging in the imaginary-time formalism.
We speculate that this geometric description has a promising potential to provide a solid basis for systematic investigation of possible anomalous transport phenomena associated with the spin current.
We leave all these interesting questions to future work.

\acknowledgments
 We thank Casey Cartwright, Markus Garbiso, Yoshimasa Hidaka, Marco Knipfer, Yu-Chen Liu, and Yuya Tanizaki for useful discussions. 
This work is supported by the U.S. Department of Energy, Office of Science, Office of Nuclear Physics under Award Number DE-FG0201ER41195, and within the framework of the Beam Energy Scan Theory (BEST) Topical Collaboration, and also partially by RIKEN iTHEMS Program (in particular iTHEMS Non-Equilibrium Working group).  X.~G.~H is supported by NSFC under Grant No.~12075061 and Shanghai NSF under Grant No.~20ZR1404100. This work is supported, in part, by the U.S. Department of Energy grant DE-SC0012447.

\appendix

\section{Global equilibrium under rotation}

\newcommand\bc{\beta_c}
\newcommand\Tc{T_c}

\label{sec:equilibrium}
A global equilibrium state is characterized by a time-like Killing vector $\beta^{\mu}$~(see, e.g., Refs.~\cite{Groot1980,Becattini:2012tc}). In the flat spacetime, it is a solution to the Killing equation
\begin{equation}
  \partial_\mu \beta_\nu + \partial_\nu \beta_\mu = 0, 
\end{equation}
and has the form 
$\beta^\mu = 
{\beta_c 
(\Omega^\mu_{~\nu} x^\nu + b^\mu)}$ with a constant anti-symmetric tensor
$\Omega_{\mu\nu} = - \Omega_{\nu\mu}$, a constant time-like vector $b^\mu$, and an overall constant factor $\beta_c$, which correspond, respectively, to angular velocity of rotation, the linear velocity of the center of rotation and the inverse temperature at the center. For instance, the fluid rigidly rotating around the $z$-axis is described by
\begin{equation}
 \beta^\mu = \beta u^\mu = \beta_c
  \begin{pmatrix}
   1 \\
   - \Omega y \\ 
   \Omega x \\
   0
  \end{pmatrix}
  \with
  \beta = \beta_c \sqrt{1 - \Omega^2 r^2}, \quad 
  u^\mu = 
  \frac{\beta_c}{\beta}
  \begin{pmatrix}
   1 \\
   - \Omega y \\
   \Omega x \\
   0 
  \end{pmatrix},
  \label{equilibriumvelocity}
\end{equation}
where $r^2=x^2+y^2$.
It can be checked explicitly that Eq.\eqref{equilibriumvelocity} is an exact solution of the ideal relativistic hydrodynamic equations. This is a consequence of the fact that conservation of angular momentum implies existence of equilibrium (nondissipative) rotating solution. Note that thermal equilibrium in the fluid requires that the local temperature $T=1/\beta$ varies in such a way that $T/\gamma$ is constant, where $\gamma=1/\sqrt{1-\Omega^2r^2}$ is the relativistic time dilation factor.

The solution given by Eq.\eqref{equilibriumvelocity} exists only for the angular velocity $\Omega$ smaller than $1/R$, where $R$ is the maximum radial size of the rotating system, i.e., $r<R$. This is obvious directly from Eq.~\eqref{equilibriumvelocity}, since $\beta$ becomes complex otherwise, and physically, since the velocity $\Omega R$ of the outer edge of the system cannot become superluminal lest it violates causality.

We show in the following that it is physically impossible to spin the fluid fast enough so that it would violate causality, since no finite amount of angular momentum could make the fluid rotate rigidly with angular velocity $\Omega$ exceeding $1/R$. The reason for that is that the temperature of the outer edge of the system 
\begin{equation}
    T(R)=\frac{1}{\beta(R)}=\frac{\Tc}{\sqrt{1-\Omega^2R^2}}
\end{equation}
approaches infinity when $\Omega\to1/R$ and, as a result, the edge carries an infinite amount of angular momentum, as we shall now show.

The orbital angular momentum along the $z$-axis is given by
\begin{equation}\label{eq:Jz}
    {\cal J}^z=\int \diff^3 r\, \left(x\,\Theta^{0y}-y\,\Theta^{0x}\right)
    =\int \diff^3 r\, wu^0\left(xu^y-yu^x\right)
    =\Omega\beta_c^2\int \diff^3r\,  {w}{\beta^{-2}}\,(x^2+y^2)\,,
\end{equation}
where $w=\epsilon + p$ is the enthalpy. For simplicity let us consider a rotating cylinder of radius $R$ and length $L$ and perform the integration in Eq.\eqref{eq:Jz} in cylindrical coordinates. Since the local temperature $T(r)=1/\beta(r)$ is a monotonous function of the radial coordinate $r=\sqrt{x^2+y^2}$, it is convenient to change the integration variables from $r$ to $T$:
\begin{equation}\label{eq:Jz-cyl}
    {\cal J}^z=2\pi L{\Omega}{\bc^2}\int_0^R \diff r r^3 w\beta^{-2} 
    = 2\pi L\Omega^{-3}\int_{\Tc}^{T(R)}
\frac{\diff T}{T}\left(1-\frac{\Tc^2}{T^2}\right) w(T)
\,.
\end{equation}
Since thermodynamic stability requires $w$ to be a non-decreasing function of $T$, the integral diverges when the edge temperature $T(R)\to\infty$, i.e., when $\Omega\to 1/R$.
 Therefore, an infinite amount of angular momentum is needed to reach the causality upper bound for the angular velocity of rigid rotation $\Omega\rightarrow 1/R$.

\section{Currents in metric-affine connection formulation}

\label{sec:metric}

The vierbein-spin connection formulation used in the main text is a useful tool to define the spin current and the canonical energy-momentum tensor. 
Since our energy-momentum tensor has a possible anti-symmetric component, one may be interested in how it appears in the formulation relying on the metric. 
We here provide a way to define the currents used in this paper with a help of the first-order metric-connection formulation~(see also, e.g., Appendix G of Ref.~\cite{Jensen:2013kka}).

Let us now use the metric $g_{\mu\nu}$ and affine connection $\Gamma^\mu_{~\nu\rho}$, instead of $e_\mu^{~\ha}$ and $\omega_{\mu~b}^{~\ha}$, as independent backgrounds, which also allows a possible contribution from the non-vanishing torsion.
While QCD enjoys diffeomorphism invariance as $\delta_\xi \Scal_{\qcd} = 0$ even if we use these backgrounds, its variation with respect to the metric and affine connection leads to a different set of currents.
We then introduce the following currents as an alternative to Eq.~\eqref{eq:Currents}:
\begin{equation}
 t^{\mu\nu} (x) 
  \equiv 
  \frac{2}{\sqrt{-g(x)}}
  \frac{\delta \Scal_{\qcd}}{\delta g_{\mu\nu}(x)} , \quad 
  s^{\mu\nu}_{~~\rho} (x) \equiv
  \frac{1}{\sqrt{-g(x)}}
  \frac{\delta \Scal_{\qcd}}{\delta \Gamma^\rho_{~\mu\nu}(x)}.
  \label{eq:Currents2}
\end{equation}
By noting Eqs.~\eqref{eq:g-e}, one can find their relations to $\Theta^{\mu}_{~\ha}$ and $\Sigma^\mu_{~\ha\hb}$ in~Eq.~\eqref{eq:Currents}.
In fact, Eqs.~\eqref{eq:g-e} allows us to express variations of the metric and affine connection as 
\begin{equation}
 \begin{split}
  \delta g_{\mu\nu} 
  &= \eta_{\ha\hb} 
  \big( 
  e_\nu^{~\hb} \delta e_\mu^{~\ha} + e_\mu^{~\ha} \delta e_\nu^{~\hb}
  \big),
  \\
  \delta \Gamma^\rho_{~\mu\nu} 
  &= 
  e_{\ha}^{~\rho} \partial_\mu \delta e_\nu^{~\ha}
  - \Gamma^\lambda_{~\mu\nu} e_{\ha}^{~\rho} \delta e_\lambda^{~\ha} 
  + \omega_{\mu~\hb}^{~\ha} e_{\ha}^{~\rho} \delta e_\nu^{~\hb}
  + e_\nu^{~\hb} e_{\ha}^{~\rho} \delta \omega_{\mu~\hb}^{~\ha}.
 \end{split}
\end{equation}
Thus, the variation of the QCD action induced by diffeomorphism is expressed as follows:
\begin{align}
  \delta_\xi \Scal_{\qcd}
  &= \int \diff^4 x
  \left[
  \frac{\delta \Scal_{\qcd}}{\delta g_{\mu\nu}} 
  \delta_\xi g_{\mu\nu}
  + \frac{\delta \Scal_{\qcd}}{\delta \Gamma^\mu_{~\nu\rho}} 
  \delta_\xi \Gamma^\mu_{~\nu\rho}
  + \frac{\delta \Scal_{\qcd}}{\delta q} 
  \delta_\xi q
  + \frac{\delta \Scal_{\qcd}}{\delta \bar{q}} 
  \delta_\xi \bar{q}
  + \frac{\delta \Scal_{\qcd}}{\delta A_\mu} 
  \delta_\xi A_\mu
  \right]
 \nonumber \\
  &= \int \diff^4 x \sqrt{-g}
  \bigg[
  \Big( t^{\mu\nu} e_{\nu \ha} 
  - (D_\lambda - \Gcal_\lambda ) s^{\lambda\mu}_{~~\ha}
  \Big)
  \delta_\xi e_\mu^{~\ha} 
  - \frac{1}{2} ( s^{\mu}_{~\ha\hb} - s^{\mu}_{~\hb\ha} )
  \delta_\xi \omega_{\mu}^{~\ha\hb}
  \bigg],
 \label{eq:metric-affine-variation}
\end{align}
from which we find the relation between the two definitions of currents,~\eqref{eq:Currents} and \eqref{eq:Currents2}, as 
\begin{equation}
  \Theta^{\mu\nu} 
   = t^{\mu\nu} - (\nabla_\lambda - \Gcal_\lambda) s^{\lambda \mu\nu }, \quad 
   \Sigma^{\mu\nu\rho} 
   = s^{\mu\nu\rho} - s^{\mu\rho\nu}. 
   \label{eq:currents-relation}
 \end{equation}
This relation allows us to investigate dynamics of 
$\Theta^{\mu\nu}$ and $\Sigma^{\mu\nu\rho}$, using 
$t^{\mu\nu}$ and $s^{\lambda\mu\nu}$ that are defined by the variation with respect to 
the metric and affine connection.

\bibliographystyle{utphys}
\bibliography{torsion-linear-response}
\end{document}